\documentclass[12pt]{article}
\usepackage{epsfig}
\usepackage{graphicx}
\usepackage{a4}
\usepackage{latexsym}
\usepackage{slashed}
\usepackage{cite}
\usepackage{scalefnt}
\usepackage{color}
\usepackage{colordvi}

\usepackage{pslatex}
\usepackage[latin1]{inputenc}
\usepackage[T1]{fontenc}

\newcommand{\as}{\alpha_{\rm s}}

\def\MSbar{\overline{\mathrm{MS}}}
\def\ep{\epsilon}
\def\z#1{{\zeta_{#1}}}
\def\ca{{C^{}_A}}
\def\cf{{C^{}_F}}
\def\tf{{T^{}_F}}
\def\nf{{n^{}_{\! f}}}

\def\lm{\mathrm{L}_m}
\def\ls{\mathrm{L}_s}
\def\lx{\mathrm{L}_x}
\def\ly{\mathrm{L}_y}
\def\s#1#2{\mathrm{S}_{#1,#2}(x)}
\def\li#1{\mathrm{Li}_{#1}(x)}

\newcommand{\brk}{\right. \nonumber \\ && \left.}
\newcommand{\ibrk}{\right. \right. \nonumber \\ && \left. \left.}
\newcommand{\dibrk}{\right. \right. \right. \nonumber \\ && \left. \left. \left.}
\newcommand{\ddibrk}{\right. \right. \right. \right. \nonumber \\ && \left.\left. \left. \left.}

\newcommand{\text}{\textstyle}
\newcommand{\bea}{\begin{eqnarray}}
\newcommand{\eea}{\end{eqnarray}}

\def\braket#1#2{\langle #1 |#2 \rangle}

\def\Ione#1{\mbox{\boldmath I}_{#1}^{(1)}(\ep)}
\def\Iee#1{\mbox{\boldmath I}_{#1}^{(1)}(2\ep)}
\def\Itwo#1{\mbox{\boldmath I}_{#1}^{(2)}(\ep)}

\def\Fintwo#1{{{\cal F}_{\!inite}^{\, (0 \times 2)}}_{#1}}

\def\Finone#1{{{\cal F}_{\!inite}^{\, (1 \times 1)}}_{#1}}

\def\nnlo{{\rm NNLO}}

\def\CA{C_A}
\def\CF{C_F}

\def\M{{\cal M}}
\def\A{{\cal A}}

\begin{document}
\setlength{\parskip}{0.2cm} \setlength{\baselineskip}{0.55cm}

\begin{titlepage}
\vspace{1.5cm}
\begin{center}
\LARGE {\bf
W Pair Production at the LHC \\[0.5ex]
I. Virtual $\mathcal{O}(\alpha_s^2)$ Corrections in the High Energy Limit
}\\
\vspace{2.2cm}
\large
G. Chachamis$^{a}$, M. Czakon$^{a,b}$ and D. Eiras$^{a}$\\
\vspace{1.4cm}
\normalsize
{\it
$^{a}$Institut f\"ur Theoretische Physik und Astrophysik, Universit\"at W\"urzburg \\[0.5ex]
Am Hubland, D-97074 W\"urzburg, Germany \\[.5cm]
$^{b}$Department of Field Theory and Particle Physics,
Institute of Physics \\[0.5ex]
University of Silesia, Uniwersytecka 4, PL-40007 Katowice,
Poland \\[.5cm]}
%
\vspace{4cm}
\large {\bf Abstract}
\vspace{-0.2cm}
\end{center}
We present the result for the two-loop and the one-loop squared
virtual QCD corrections to the W boson pair production in the 
quark-anti-quark-annihilation channel in the limit where all 
kinematical invariants are large compared to the mass of the W boson.
The infrared pole structure is in agreement with the
prediction of Catani's general formalism for the singularities of
two loop amplitudes.
\\
\vspace{2.0cm}
\end{titlepage}

\newpage

%
%
\section{Introduction}
\label{sec:intro}
The Large Hadron Collider (LHC) will be
the centre of interest for particle physics phenomenology in the next years.
Open issues that require definite answers are the 
verification of the consistency and
validity of the Standard Model (SM) in the energy range of the LHC
as well as insights into New Physics. 
Several proposed models and concepts that have the SM as their low energy
limit theory are either to pass the LHC test 
or to be proven wrong.
Supersymmetry and Extra-dimensions are two of the most illustrious 
examples.

Probably, the most important goal for the
LHC is the discovery of the elusive Higgs boson. The latter
is part of the mechanism of dynamical breaking of
the Electroweak (EW) symmetry and is responsible for the fermions
and gauge bosons mass. Discovering the only constituent of the 
Standard Model (SM) which has not been experimentally observed yet,
along with a systematic measurement of its properties, will be essential
for our understanding of mass and the precise gauge structure of the SM.
Another important endeavour at the LHC, in connection to the investigation
of the non-Abelian gauge structure of the SM, is the precise measurement of the
hadronic production of gauge boson pairs, 
$W W$, $W Z$, $Z Z$, $W \gamma$, $Z \gamma$.
Deviations from the SM predictions would indicate the presence of either
anomalous couplings or new heavy particles which would decay into
vector boson pairs~\cite{tevatron1,tevatron2}.

Seen under the prism of the previous argumentation,
W pair production via quark-anti-quark-annihilation,
\begin{equation}
q {\bar q} \rightarrow W^+ \, W^- \, ,
\end{equation}
is a very important process at the LHC.
Firstly, it can serve as a signal process 
in the search for New Physics since it can be used
to measure the vector boson trilinear couplings as predicted by the
Standard Model (SM) (actually, this is the favored channel as it
involves both trilinear vertices, $W W Z$ and $W W \gamma$).
Secondly,  $ q {\bar q} \rightarrow W^+ W^-$ is the dominant irreducible
background to the promising Higgs discovery channel 
\begin{equation}
p p \rightarrow H \rightarrow W^* W^* 
\rightarrow l {\bar \nu} {\bar l}' \nu' \, ,
\end{equation}
in the mass range M$_{\mathrm{Higgs}}$ 
between 140 and 180 GeV~\cite{dittmardreiner}.

Due to its importance, the study of W pair production in hadronic 
collisions has
attracted a lot of attention in the literature. The Born cross section was
calculated almost 30 years ago~\cite{brown}, whereas the 
next-to-leading order (NLO)
QCD corrections to the tree-level were computed 
in Refs.~\cite{ohn,fri,dixon1,dixon2,campbell}
and were proven to be large. 
They enhance the tree-level by
almost 70\% which falls to a (still) large 30\% after 
imposing a jet veto. 
Therefore, if a theoretical estimate for  the
W pair production is to be compared against 
experimental measurements at the LHC, one is bound to 
go one order higher in the perturbative expansion, namely
to the next-to-next-to-leading order (NNLO). This would 
allow, in principle, an accuracy of better than 10\%.
Notice that first steps in this direction have been done
by considering soft-gluon resummation effects in W pair 
production~\cite{Grazzini:2005vw}.

High accuracy for the W pair production is also needed
when the process is studied as background to Higgs production.
The NLO QCD corrections to the signal process for the Higgs discovery via
gluon fusion,
$g g \rightarrow H$, 
contribute a 70\%~\cite{Spira:1995rr,Dawson:1990zj}, whereas
the NNLO contributions suggest an additional 20\% for the 
LHC~\cite{Harlander:2002wh,Anastasiou:2002yz,Ravindran:2003um}. 
With a jet veto, at NNLO the 
total corrections are of the order of 
85\%~\cite{Catani:2001cr,Davatz:2004zg,Anastasiou:2004xq}.
Lastly, the QCD corrections to the cross section for the process
$H \rightarrow W W \rightarrow l {\bar \nu} {\bar l}' \nu'$ 
are known at NNLO~\cite{Anastasiou:2007mz,
Grazzini:2008tf} whereas the EW ones are known
beyond NLO~\cite{Bredenstein:2006rh}.
The ratio of the Higgs signal over background 
is expected between 1:1 and 2:1 once certain cuts are applied that
reject events with high $p_T$ jets. For a consistent QCD analysis,
therefore, we need to compare both signal and background cross sections
calculated at the same order, that is, at NNLO.
Another process that needs to be included in the background
is the W pair production
in the loop induced gluon fusion channel, 
\begin{equation}
g g \rightarrow W^+ W^- \, .
\end{equation}
This 
contributes at $\mathcal{O}(\alpha_s^2)$ relative to the 
quark-anti-quark-annihilation channel but is 
nevertheless enhanced due to the large gluon flux
at the LHC. The corrections from gluon fusion increase the W pair background
estimate by almost  30\% after certain experimental Higgs search cuts 
are imposed~\cite{kauer1,kauer2}.

In this paper, we address the task
of computing the NNLO virtual part, more precisely
the interference of the two-loop with the Born amplitude, as well as
the the one-loop squared contribution.
We work in the limit of fixed scattering angle and high energy, where all
kinematical invariants are large compared to the mass $m$ of the W.
Our result contains all logarithms $\log m$ as well as all constant
contributions while we neglect power corrections in $m$. These will be
presented in a following publication.

Our methodology for obtaining
the massive amplitude (massless fermion-boson scattering was studied 
in Ref.~\cite{Anastasiou:2002zn}) 
is very similar to the one followed 
in Refs.~\cite{qqTT,ggTT,Czakon:2004wm} which is, at its turn, an evolution
of the methods employed in Refs.~\cite{Czakon:2006pa,Actis:2007gi}.
The amplitude is reduced to an expression that only
contains a small number of integrals (master integrals)
with  the help of the Laporta algorithm~\cite{Laporta:2001dd}.
In the calculation for the two-loop amplitude there are 71 master integrals.
For the one-loop squared case, we use the helicity matrix formalism
to reduce the problem to a small set of integrals.
Next comes the construction, in a fully automatised way,
of the Mellin-Barnes (MB) 
representations~\cite{Smirnov:1999gc,Tausk:1999vh}
of all the master integrals by using
the {\tt MBrepresentation} package~\cite{MBrepresentation}. The
representations  are then
analytically continued in the number of space-time dimensions by means of the
{\tt MB} package~\cite{Czakon:2005rk}, thus revealing the full singularity
structure. An asymptotic 
expansion in the mass parameter is performed by
closing contours and the integrals are finally resummed,
either with the help of {\tt XSummer}~\cite{Moch:2005uc}
or the {\tt PSLQ} algorithm~\cite{pslq:1992}.

Our paper is organised as follows. In Section~\ref{sec:notation} we
introduce our notation, present briefly our methods 
and define the perturbative expansion of the
matrix elements summed over colours and spins.  
In Section~\ref{sec:catani} we study the singular behavior of  the NNLO
contributions, and verify that it agrees with the general formalism 
developed by Catani~\cite{catani} for the infrared structure of QCD
amplitudes.  In
Section~\ref{sec:results} we present the finite remainders for the
interference of the tree and the two-loop amplitude and the 
one-loop squared after subtraction of
the singular poles of Section~\ref{sec:catani} from the explicit  result. 
We organise the finite part  according
to  the colour content of the two-loop amplitude for the two-loop case.  
The finite remainders are 
expressed in terms of logarithms and polylogarithms which are real in
the  physical domain. We conclude in Section~\ref{sec:conclusions}.
Finally, for completeness, the one-loop result
up to order $\epsilon^2$ is included in the Appendix.

%
%
\section{Notation}
\label{sec:notation}
The charged vector-boson 
production in the leading partonic scattering process
corresponds to
\begin{equation}
\label{eq:qqWW}
q_j(p_1) + {\overline q}_j(p_2) 
\:\:\rightarrow\:\: W^-(p_3,m) + W^+(p_4,m) \, ,
\end{equation}
where $p_i$ denote 
the quark and W momenta, $m$ is the mass of the W boson and
j is a flavour index.
We are considering down type quark scattering in our paper. Obtaining
the corresponding result for up-type quark scattering is actually trivial as
we will show in the following.
Energy-momentum conservation implies
\begin{equation}
\label{eq:engmom}
p_1^\mu+p_2^\mu = p_3^\mu+p_4^\mu \, .
\end{equation}
We consider
the scattering amplitude ${\cal M}$ for the process~(\ref{eq:qqWW})
at fixed values of the external parton momenta $p_i$, thus $p_1^2 =
p_2^2 = 0$ and $p_3^2 = p_4^2 = m^2$.
The amplitude ${\cal M}$ may be written as a 
series expansion in the strong coupling $\as$,
\begin{eqnarray}
  \label{eq:Mexp}
  | {\cal M} \rangle
  & = &
  \biggl[
  | {\cal M}^{(0)} \rangle
  + \biggl( {\as \over 2 \pi} \biggr) | {\cal M}^{(1)} \rangle
  + \biggl( {\as \over 2 \pi} \biggr)^2 | {\cal M}^{(2)} \rangle
  + {\cal O}(\as^3)
  \biggr]
\, ,
\end{eqnarray}
and we define the expansion parameter in powers of $\as(\mu^2)
/ (2\pi)$ with $\mu$ being the renormalisation scale. We work in
conventional dimensional regularisation, $d=4-2 \ep$, in the
$\MSbar$-scheme for the coupling constant renormalisation.

We explicitly relate the bare (unrenormalised) coupling $\as^{\rm{b}}$
to the renormalised coupling $\as$ by
\begin{eqnarray}
\label{eq:alpha-s-renorm}
\as^{\rm{b}} S_\epsilon \: = \: \as
\biggl[
   1
   - {\beta_0 \over \epsilon} \biggl( {\as \over 2 \pi} \biggr)
  + {\cal O}(\as^2)
  \biggr]
\, ,
\end{eqnarray}
where we set the factor 
$S_\epsilon=(4 \pi)^\ep \exp(-\ep \,\gamma_{\rm E}) = 1$
for simplicity and $\beta$ is the QCD $\beta$-function known at present up to
the four-loop level
\cite{vanRitbergen:1997va,Czakon:2004bu}
\begin{eqnarray}
\label{eq:betafct}
\beta_0 = {11 \over 6}\*\ca - {2 \over 3}\*\tf\*\nf \, .
\end{eqnarray}
The color factors in a non-Abelian ${\rm{SU}}(N)$-gauge theory are
$\ca = N$, $\cf = (N^2-1)/2\*N$ and $\tf = 1/2$.
Throughout this paper, $N$ denotes the number of colors and
$\nf$ the total number of flavors of massless quarks. Remark, however, that
the latter must come in pairs, because of the flavor changing coupling to the
charged gauge boson. This is only problematic in the case of top quarks
running in a closed loop.

In the following, our discussion will be restricted to
the two-loop amplitude summed over spins and colours and
contracted with the Born one.
Nevertheless, it should be stressed that our methods
and the results 
of the present work can be easily extended to the
partial amplitudes for the individual helicity 
combinations of the massive two-loop amplitude
$|{\cal M}^{(2)} \rangle$ itself.

For convenience, we define the function ${\cal A}(\epsilon, m, s, t, \mu)$
for the squared amplitudes summed over spins and colors as
\begin{eqnarray}
\label{eq:Msqrd}
\overline{\sum |{\cal M}({q_j + {\overline q}_j \to  W^+ + W^-} )|^2}
&=&
{\cal A}(\epsilon, m, s, t, \mu)
\, .
\end{eqnarray}
${\cal A}$ is a function of the Mandelstam variables $s$, $t$ and $u$ given by
\begin{equation}
\label{eq:Mandelstam}
s = (p_1+p_2)^2\, , \qquad
t  = (p_1-p_3)^2 - m^2\, , \qquad
u  = (p_1-p_4)^2 - m^2\, ,
\end{equation}
and has a perturbative expansion similar to Eq.~(\ref{eq:Mexp}),
\begin{equation}
\label{eq:Aexp}
{\cal A}(\epsilon, m, s, t, \mu) = 
\left[
  {\cal A}^{(0)}
  + \biggl( {\as \over 2 \pi} \biggr) {\cal A}^{(1)}
  + \biggl( {\as \over 2 \pi} \biggr)^2 {\cal A}^{(2)}
  + {\cal O}(\as^{3})
\right]
\, .
\end{equation}
In terms of the amplitudes the expansion coefficients in Eq.~(\ref{eq:Aexp})
may be expressed as
\begin{eqnarray}
\label{eq:A4def}
{\cal A}^{(0)} &=&
\langle {\cal M}^{(0)} | {\cal M}^{(0)}\rangle \, , \\
\label{eq:A6def}
{\cal A}^{(1)} &=& \left(
\langle {\cal M}^{(0)} | {\cal M}^{(1)} \rangle + \langle {\cal M}^{(1)} | {\cal M}^{(0)} \rangle
\right)\, , \\
\label{eq:A8def}
{\cal A}^{(2)} &=& \left(
\langle {\cal M}^{(1)} | {\cal M}^{(1)} \rangle
+ \langle {\cal M}^{(0)} | {\cal M}^{(2)} \rangle + \langle {\cal M}^{(2)} | {\cal M}^{(0)} \rangle
\right)\, ,
\end{eqnarray}
where ${\cal M}^{(0)}$ and ${\cal M}^{(1)}$ are the massive tree level
and one loop amplitudes correspondingly.
${\cal A}^{(0)}$ is given by
\begin{eqnarray}
{\cal A}^{(0)} &=& N \left \{ 
c_1 \left[ 16 (1- \epsilon)^2 \frac{x}{\left (1-x \right )} 
+ 4 (3 - 4\epsilon) \frac{1}{m_s} 
+ \frac{4 x\,(1-x)}{m_s^2}\right ] 
\brk
+ c_2 \left [ -24 + 16 x + 16 \epsilon 
\left ( 2 - x \right ) + 4 \frac{(3 - 4 \epsilon) 
- 2 x (1-x)}{m_s} +  \frac{4 x \,(1-x)}{m_s^2}\right ] 
\brk
+ c_3 \left [ - 24 \left ( 1-x\, (1-x)\right ) 
+ 16 \epsilon \left ( 2 - x \, (1-x)\right ) 
+ \frac{6 - 8 \epsilon - 8 x \, (1-x)}{m_s} 
+ \frac{2 x \, (1-x)}{m_s^2} \right ] \right \} \, ,
   \nonumber \\ 
\end{eqnarray}
where we have defined $x = -\frac{t}{s}$, $m_s = \frac{m^2}{s}$ and only the
leading physical powers (i.e. down to the constant) in the $m_s$-expansion are
retained. Notice that, once the actual values of the $c_i$ are substituted, the
terms singular in $m_s$ cancel as required by unitarity.
This will be the case  for the
final two-loop and one-loop squared expressions as well. 
The coefficients $c_1$,
$c_2$ and $c_3$ are in their essence
combinations of EW coupling constants defined as
\begin{eqnarray}
c_1 &=& \frac{g_{WL}^4}{4} \, , \nonumber \\
c_2 &=& \frac{1}{4 \, s_w^2}\left(Q_q+2 g_{ZL}^q \frac{c_w }{s_w \left (
      1-\frac{M_Z^2}{s}\right )}\right )
\, , \nonumber \\
c_3 &=& \frac{c_w^2}{s_w^2 (1-\frac{M_Z^2}{s})^2} \left (  ( g_{ZA}^q)^2 +
  \left (g_{ZV}^q + Q_q \frac{s_w \left ( 1-\frac{M_Z^2}{s} \right )}{c_w} 
\right )^2 \right ) \, .
\end{eqnarray}

In order to compute the $\langle {\cal M}^{(1)} | {\cal M}^{(1)}
\rangle$ it proves convenient to express the 
amplitude $| {\cal M}^{(1)} \rangle$  in terms of helicity
amplitudes, ${\mathcal M}^g(\lambda_1, \lambda_2, s, t)$,
where $\lambda_1$ and $\lambda_2$ stand for the
helicities of the W$^+$ and W$^-$ respectively.
In other words, it is convinient to  
decompose the amplitude
into a sum of products consisting generally of three parts,
a Feynman integral, 
a rational function of the kinematical variables and a 
standard  matrix element, ${\mathcal M_j^g}$, 
a complete list of which is listed below in Eq.~(\ref{eq:standardHME}).
For that, Dirac algebra is used, as well as the equations of motion and
an anticommuting $\gamma_5$.
The quark and anti-quark have opposite helicities in the
centre-of-mass system so one helicity label above, $g = \pm 1$,
suffices. 

Therefore, the one-loop amplitude is formally
rearranged as 
\bea
\label{eq:hme}
| {\cal M}^{(1)} \rangle = \sum_{i,j,g}  C_i (s,t,u) 
{\mathcal I}_i^j (s,t,u;\mu^2) {\mathcal M_j}(\{p_k\},g)\, , 
\eea
where the $C_i$ are coefficients, the ${\mathcal I}_i^j$ are one-loop
dimensionally regularized scalar
integrals, ${\mathcal M_j}$ are helicity matrix elements, g = $\pm$ and $k =
1,...,4$. 
The ten helicity matrix elements ${\mathcal M_j}(p_k,g) = 
{\mathcal  M_j^g}$ have been taken as defined in Ref.~\cite{Diener:1997nx} 
(see also~\cite{Denner:1988tv}):
\bea
\label{eq:standardHME}
{\mathcal M}_0^g &=& {\overline v }(p_2) \, \slashed{\varepsilon}_1 (\slashed{p}_3 -
\slashed{p}_2) 
\slashed {\varepsilon}_2
{\mathcal P}_g \, u(p_1)  \, , \nonumber \\
{\mathcal M}_1^g &=& {\overline v}(p_2) \, \slashed{p}_3 {\mathcal P}_g \, u(p_1) \,
\varepsilon_1 \cdot \varepsilon_2 \, , \nonumber \\
{\mathcal M}_2^g &=& {\overline v}(p_2)\, \slashed{\varepsilon}_1 {\mathcal P}_g \, u(p_1)
\, \varepsilon_2 \cdot p_3 \, , \nonumber \\
{\mathcal M}_3^g &=& -{\overline v}(p_2)\,\slashed{\varepsilon}_2 {\mathcal P}_g \, u(p_1)
\, \varepsilon_1 \cdot p_4 \, , \nonumber \\
{\mathcal M}_4^g &=& {\overline v}(p_2)\, \slashed{\varepsilon}_1 {\mathcal P}_g \, u(p_1)
\, \varepsilon_2 \cdot p_1\, , \nonumber \\
{\mathcal M}_5^g &=& -{\overline v}(p_2)\, \slashed{\varepsilon}_2 {\mathcal P}_g \, u(p_1)
\, \varepsilon_1 \cdot p_2\, ,  \\
{\mathcal M}_6^g &=& {\overline v}(p_2)\, \slashed{p}_3 {\mathcal P}_g \, u(p_1) \,
\varepsilon_1 \cdot p_2 \, \varepsilon_2 \cdot p_1
\, , \nonumber \\
{\mathcal M}_7^g &=& {\overline v}(p_2)\, \slashed{p}_3 {\mathcal P}_g \, u(p_1) \,
\varepsilon_1 \cdot p_2 \, \varepsilon_2 \cdot p_3 \, , \nonumber \\
{\mathcal M}_8^g &=& {\overline v}(p_2)\, \slashed{p}_3 {\mathcal P}_g \, u(p_1)\,
\varepsilon_1 \cdot p_4 \, \varepsilon_2 \cdot p_1 \, , \nonumber \\
{\mathcal M}_9^g &=& {\overline v}(p_2)\, \slashed{p}_3 {\mathcal P}_g \, u(p_1)\,
\varepsilon_1 \cdot p_4 \, \varepsilon_2 \cdot p_3 \, , \nonumber
\eea
where ${\mathcal P}_g = {\mathcal P}_{\pm} = \frac{1 \pm \gamma_5}{2}$.
All colour indices as well 
as the arguments of the polarization vectors,
$\varepsilon_1(p_3, \lambda_1)$ and
$\varepsilon_2(p_4, \lambda_2)$, have been suppressed.

After expressing the one-loop amplitude as in Eq.~(\ref{eq:hme}) and 
calculating the
Feynman integrals, it is trivial to obtain  
$\langle {\cal M}^{(1)} | {\cal M}^{(1)}
\rangle$, one needs only to 
compute the traces (one fermionic chain in all cases)
coming from multipling a matrix element with the 
complex congugate of another one.
We have decomposed the tree level amplitude 
as well in terms of helicity amplitudes
and computed $\langle {\cal M}^{(0)} | {\cal M}^{(1)}
\rangle$ as a trivial cross check.
Even though the representations in Eq.~(\ref{eq:standardHME}) have been 
used internally,
here we present our result only for the amplitude squared and summed over 
helicities. Notice that this is done in conventional dimensional
regularization, which implies $2-2\ep$ polarizations of the vector bosons.

The expressions for ${\cal A}^{(1)}$ have been presented e.g. in 
Refs.~\cite{ohn, fri} whereas the leading color coefficient of 
$\langle {\cal M}^{(0)} | {\cal M}^{(2)} \rangle$ was discussed
in Ref.~\cite{Chachamis:2007cy}.
Here we provide for the first time the result for the real part of
${\cal A}^{(2)}$.

%
%
%
%
\section{Infrared Pole Structure}
\label{sec:catani}

In the simpler case of one-loop  amplitudes,
their poles in $\ep$ can be  
expressed as a universal combination of the tree amplitude 
and a colour-charge operator $\Ione{}$. 
The generic form of the $\Ione{}$ operator was found by Catani and 
Seymour~\cite{Catani:1996vz} (see also~\cite{Giele:1991vf,Kunszt:1994mc})
and it was derived for the general one-loop QCD 
amplitude by integrating the real radiation graphs of the same order  in
perturbation series in the one-particle unresolved limit.

The pole structure of our one-loop expression is given, 
according to the prediction by Catani, by acting with the operator
$\Ione{}$ onto the tree-level result:
\bea
| {\cal M}^{(1)} \rangle &=& \Ione{}  | {\cal M}^{(0)} \rangle + | {\cal
  M}^{(1)}_{\rm finite} \rangle \, ,
\eea
where $\Ione{}$ is defined as
\begin{eqnarray}
\Ione{} = -\CF \frac{e^{\ep \gamma}}{\Gamma(1-\ep)} 
\left(\frac{1}{\ep^2} + \frac{3}{2\ep} \right) 
\left(-\frac{\mu^2}{s}\right)^\ep\, .
\label{eq:Ioneqq}
\end{eqnarray}

In a similar way, the divergences of the
two-loop amplitude can be  written as a sum of two terms: the action of
the $\Ione{}$ operator on the one-loop amplitude and the action of a new
operator $\Itwo{}$ on the tree  amplitude. The $\Itwo{}$ operator
includes a renormalisation scheme dependent term $H^{(2)}$ multiplied by
a $1/\ep$ pole. 
In the following, we give explicit
expressions for $\Ione{}$ and $\Itwo{}$ which are
valid in the $\MSbar$ scheme.

At next-to-next-to-leading-order (NNLO), contributions from 
the self-interference of the one-loop amplitude and the interference of 
the tree and the two-loop amplitude must be taken into account, so that
\begin{equation}
\A ^\nnlo (s, t, u, m, \mu) =\A^{\nnlo (1 \times 1)}(s, t, u, m, \mu)
+ \A^{\nnlo (0 \times 2)}(s, t, u, m, \mu),
\end{equation}
with 
\begin{equation}
\A^{\nnlo (1 \times 1)}(s, t, u, m, \mu)=\braket{\M^{(1)}}{\M^{(1)}},
\end{equation}
and 
\begin{equation}
\A^{\nnlo (0 \times 2)}(s, t, u, m, \mu)=\braket{\M^{(0)}}{\M^{(2)}}
  +\braket{\M^{(2)}}{\M^{(0)}}.
\end{equation}

We further decompose the one-loop self-interference and the 
two-loop contributions as a sum of singular and finite terms, 
\begin{equation}
\A^{\nnlo \, (1\times 1)}(s, t, u, m, \mu)
= \mathcal{C}_{atani}^{(1 \times 1)}(s, t, u, m, \mu) 
+ \Finone{}(s, t, u, m, \mu)
\end{equation} 
and
\begin{equation}
\A^{\nnlo \, (0\times 2)}(s, t, u, m, \mu)
 = \mathcal{C}_{atani}^{(0 \times 2)}(s, t, u, m, \mu) 
+ \Fintwo{}(s, t, u, m, \mu),
\end{equation} 

$\mathcal{C}_{atani}^{(1 \times 1)}$ and $\mathcal{C}_{atani}^{(0 \times 2)}$
contain infrared singularities that 
will be analytically canceled by the infrared 
singularities occurring in radiative processes of the same order 
(ultraviolet divergences having already been removed by 
renormalisation). $\Finone{}$ and $\Fintwo{}$ are the remainders 
which are finite as $\ep \to 0$.

The infrared poles of the interference of the tree and the two-loop
amplitudes follow a generic formula developed by Catani in Ref.~\cite{catani}.
Due to the simple colour structure of the process~(\ref{eq:qqWW})
the action of 
$\Ione{}$ and $\Itwo{}$ is factorised such that we formally have 
\begin{eqnarray}
\label{eq:fini}
\mathcal{C}_{atani}^{(1 \times 1)}(s, t, u, m, \mu) = 
|\Ione{}|^2 \langle {\cal
  M}^{(0)} | {\cal M}^{(0)} \rangle  
+ 2  {\rm Re} \left \{ \Ione{}^{*}  \langle {\cal
  M}^{(0)} | {\cal M}^{(1)}_{\rm finite} \rangle \right \} \,.
\end{eqnarray} 
and
\begin{equation}
\mathcal{C}_{atani}^{(0 \times 2)}(s, t, u, m, \mu)=2 {\rm Re} \left\{ 
\Ione{} \braket{M^{(0)}}{M^{(1)}} 
+ \Itwo{} \braket{M^{(0)}}{M^{(0)}} \right\} 
\end{equation} 
with
\begin{eqnarray}
\Itwo{} &=& -\frac{1}{2} \Ione{} \left(\Ione{}
+\frac{2\beta_0}{\ep} \right)
+\frac{e^{-\ep \gamma} \Gamma(1-2\ep)} {\Gamma(1-\ep)}
\left(\frac{\beta_0}{\ep}+ K \right) \Iee{} \nonumber \\
&& + H^{(2)}(\ep) \, ,
\end{eqnarray}
where
\begin{equation}
K=\left(\frac{67}{18} -\frac{\pi^2}{6}\right) \CA -\frac{10}{9} \tf \nf.
\end{equation}
The renormalisation scheme dependent $H^{(2)}$ constant for a QCD
amplitude with a $q \bar q$ pair is given by
\begin{eqnarray}
H^{(2)}(\ep) &=&2 \frac{e^{\ep \gamma}}{4 \ep \Gamma(1-\ep)} 
\left(-\frac{\mu^2}{s}\right)^{2 \ep} 
\left \{
\left(\frac{\pi^2}{2}-6 ~\zeta_3 
-\frac{3}{8}\right) \CF^2
\brk
+ \left(\frac{13}{2}\zeta_3 
+ \frac{245}{216}-\frac{23}{48} \pi^2 \right) \CA \CF
+  \left(-\frac{25}{54}+\frac{\pi^2}{12} \right) \CF \tf \nf 
\right \}\, .
\end{eqnarray}

We were able to verify that our results have the same infrared
structure as the one predicted by Catani's formalism.


%
%
\section{Results}
\label{sec:results}

\subsection{Two-loop Contribution}

In this section, we give explicit expressions for the finite remainder of 
the two-loop contribution $\Fintwo{}$ defined as 
\begin{equation}
 \Fintwo{}(s, t, u, m, \mu) = \A^{\nnlo \, (0\times 2)}(s, t, u, m, \mu) 
     - \mathcal{C}_{atani}^{(0 \times 2)}(s, t, u, m, \mu)\, ,
\end{equation} 
or in the rescaled form
\begin{equation}
\Fintwo{}(m_s, x, \frac{s}{\mu^2}) = 
\A^{\nnlo \, (0\times 2)}(m_s, x,\frac{s}{\mu^2} ) 
- \mathcal{C}_{atani}^{(0 \times 2)}(m_s, x, \frac{s}{\mu^2})\, .
\end{equation} 

The EW structure of the finite remainder for a down-type quark
can be factorised as 
\begin{eqnarray}
\label{eq:down2}
{\cal F}_{\!inite, \,\, down}^{\, (0 \times 2)} &=& 
2\, N \, \sum_{i=1, 4}\, c^{}_{i}  
{\cal J}_{i,\,down}^{(1 \times 1)}(m_s, x, \frac{s}{\mu^2})  \, .
\end{eqnarray}
This decomposition allows one 
to easily obtain the result for the up-type quark
scattering. The latter is then given by
\begin{eqnarray}
\label{eq:up}
{\cal F}_{\!inite, \,\, up}^{\, (0 \times 2)} &=& 
2\, N \, \sum_{i=1, 4}\, c^{}_{i}  
{\cal J}_{i,\,up}^{(0 \times 2)}(m_s, x, \frac{s}{\mu^2})  \, ,
\end{eqnarray}
where one needs to use the following formulae
\begin{equation}
{\cal J}_{1,\,up}^{(0 \times 2)}(m_s, x, \frac{s}{\mu^2})\,=
\,{\cal J}_{1,\,down}^{(0 \times 2)}(m_s, y, \frac{s}{\mu^2})\, ,
\end{equation}
\begin{equation}
{\cal J}_{2,\,up}^{(0 \times 2)}(m_s, x, \frac{s}{\mu^2})\,=
\,- {\cal J}_{2,\,down}^{(0 \times 2)}(m_s, y, \frac{s}{\mu^2})\, ,
\end{equation}
\begin{equation}
{\cal J}_{3,\,up}^{(0 \times 2)}(m_s, x, \frac{s}{\mu^2})\,=
\,{\cal J}_{3,\,down}^{(0 \times 2)}(m_s, y, \frac{s}{\mu^2})\, , 
\end{equation}
\begin{equation}
{\cal J}_{4,\,up}^{(0 \times 2)}(m_s, x, \frac{s}{\mu^2})\,=
\,{\cal J}_{4,\,down}^{(0 \times 2)}(m_s, y, \frac{s}{\mu^2})\, 
\end{equation}
and naturally to make the corresponding changes in the definitions of
the couplings $c_1$, $c_2$, $c_3$ and $c_4$, namely to 
use the up-type quark charge
and isospin. Here $y\,=\,-\frac{u}{s}$. In the following and
with no loss of clarity, since our result assumes down-type 
quark scattering, we will
suppress all indices that indicate the type of scattered quark.
The functions ${\cal J}_i(m_s, x, \frac{s}{\mu^2})$ 
in Eq.~(\ref{eq:down2}) will be 
presented decomposed according to
the colour structure, namely in the form
\begin{eqnarray}
{\cal J}_{i}^{(0 \times 2)}(m_s, x, \frac{s}{\mu^2}) &=& \left ( j^{(1)}_{i} C_F C_A 
+ j^{(2)}_{i} C_F^2 +  j^{(3)}_{i} \cf \tf \nf \right)\, .
\end{eqnarray}
$c_4$, in addition to $c_1$, $c_2$ and $c_3$,
is a new coupling that appears at the two-loop level
and is defined as:
\begin{eqnarray}
c_4 &=& - \frac{c_w \, g_{ZA}^q}{2 \, 
s_w^3 (1-\frac{M_Z^2}{s})} \, . 
\end{eqnarray}

The appearance of $c_4$ is an effect that comes from a specific part of
$\langle {\cal M}^{(0)} | {\cal M}^{(2)} \rangle$.
This part consists of two-loop fermionic boxes contracted with
the Born diagram that involves an $s$-channel $Z$ exchange.
A typical example can be seen in Fig.~(1).
The main feature of these 
diagrams is that their EW couplings fall into two disjoint
fermionic chains and once the traces are computed the axial part
drops out. By adding and subtracting to the surviving
vector part the corresponding axial part, one can combine vector
and axial contributions into a piece proportional to $c_2$. The remaining piece
is proportional to what we have defined as $c_4$.

\begin{center}
\epsfig{file=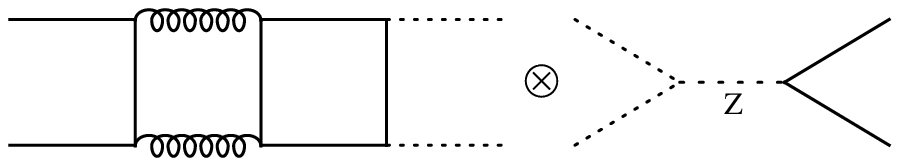,width=12cm}\\
\end{center}
{\it Fig 1}: Born diagram with a $Z$ exchanged in the $s$-channel
contracted with a fermionic two-loop box.\\

We have verified that applying
the naive recipe of sending all traces that contain a single $\gamma_5$
independently to zero is a valid approach for this class of diagrams.
We did this
by calculating explicitly the output after substituting $\gamma_5$ by its
alternative form 
$\gamma_5 = \frac{i}{4!} \epsilon_{\mu \nu \alpha \beta} 
\gamma_{\mu}\gamma_{\nu}\gamma_{\alpha}\gamma_{\beta}$
and confirming that no additional terms survive.
This was in fact a non-trivial cancellation 
as it occurs only for the sum of all the diagrams of this particular class.
Note, however, that there were finite contributions from traces containing
$\gamma_5$ in the case of pure W boson pair 
exchange (no photons or Z's involved).

We are finally ready to present our result.
The functions ${\cal J}_i$, are given by

\footnotesize
\bea
\label{eq:theend}
&&{\cal  J}_1^{(0 \times 2)} (m_s,x) =  \nonumber \\
&& \ca \cf \left \{
\frac{1}{\text{m_s}^2}
\left[
\frac{31}{120} (1-x) x \pi^4-\frac{107}{36} (1-x) x
   \pi^2-\frac{51157}{648} (1-x) x+\frac{659}{18} (1-x) x
   \z3+\frac{88}{3} (1-x) x \ls
\right]
\brk
+
\frac{1}{\text{m_s}}
\left[
\frac{31 \pi^4}{40}-\frac{107
   \pi^2}{12}+\frac{659 \z3}{6}+88
   \ls-\frac{51157}{216}
\right]
+
\left[
\frac{1}{30} \left(-684 x^3+684 x^2-114 x+\frac{31}{1-x}-31\right)
   \pi^4
\ibrk
+\frac{1}{9} \left(1404 x^3-1404 x^2+188
   x-\frac{303}{1-x}+359-\frac{108}{x}\right) \pi^2-\frac{8}{3}
   \left(2-\frac{1}{1-x}\right) \li2
   \pi^2+\frac{1}{3} \left(6 x^3-6 x^2+x\right)
   \lx^4
\ibrk
+\frac{1}{3} \left(6 x^3-6 x^2+x-2\right)
   \ly^4-\frac{4}{3} \left(36 x^3-30 x^2+11
   x-\frac{3}{1-x}+3\right) \lx^3-\frac{8}{3} \left(6 x^3-6
   x^2+x\right) \lm \lx^3
\ibrk
+\frac{4}{9} \left(-108
   x^3+126 x^2-33 x-\frac{22}{1-x}+43-\frac{15}{x}+\frac{9}{x^2}\right)
   \ly^3-\frac{8}{3} \left(6 x^3-6 x^2+x\right)
   \lm \ly^3
\ibrk
+\frac{4}{3} \left(6 x^3-6
   x^2+x+2\right) \lx \ly^3+\left(4 \left(6 x^3-6
   x^2+x\right) \pi^2+24 x\right) \lm^2+4 \left(6
   x^3-6 x^2+x\right) \lm^2 \lx^2
\ibrk
+\left(4 \left(39
   x^3-15 x^2+x-\frac{4}{1-x}+7\right)-\frac{14}{3} \left(6 x^3-6 x^2+x\right)
   \pi^2\right) \lx^2-8 \left(6 x^3-6
   x^2+x\right) \li2 \lx^2
\ibrk
+4 \left(36 x^3-24 x^2+7
   x-\frac{3}{1-x}+3\right) \lm \lx^2+4 \left(6
   x^3-6 x^2+x\right) \lm^2 \ly^2-2 \left(6 x^3-6
   x^2+x+\frac{2}{1-x}\right) \lx^2
   \ly^2
\ibrk
+\left(-\frac{2}{3} \left(42 x^3-42 x^2+7
   x+\frac{2}{1-x}+2\right) \pi^2-\frac{2}{9} \left(-702
   x^3+1134 x^2-378
   x-\frac{167}{1-x}+185-\frac{144}{x}+\frac{108}{x^2}\right)\right)
   \ly^2
\ibrk
+4 \left(36 x^3-48 x^2+15
   x-5+\frac{5}{x}-\frac{3}{x^2}\right) \lm
   \ly^2+\frac{44}{3} \left(1-\frac{1}{1-x}\right)
   \ls \ly^2+8 \left(3 x^2-x\right)
   \lx \ly^2
\ibrk
+\frac{1}{162} \left(-15480
   x+\left(-46656 x^3+46656 x^2-9072 x+\frac{19836}{1-x}-15948\right)
   \z3-\frac{58897}{1-x}+58897\right)
\ibrk
-8 \left(36 x^3-36 x^2+15
   x-\frac{4}{1-x}+5\right) \li3-16 \left(6 x^3-6 x^2+x\right)
   \li4+\left(4 \left(36 x^3-36 x^2+11 x-1\right)
   \pi^2
\dibrk
-\frac{164 x}{9}+\left(-96 x^3+96 x^2-16 x\right)
   \z3+\frac{596}{9 (1-x)}-\frac{1888}{9}\right) \lm-16
   \left(6 x^3-6 x^2+x\right) \li3
   \lm
\ibrk
-\frac{44}{3} \left(-2 x-\frac{9}{1-x}+9\right)
   \ls+\frac{88}{3} \left(x-\frac{1}{1-x}+2\right)
   \lm \ls-16 \left(3 x^2-2 x\right)
   \lm^2 \lx
\ibrk
+\left(8 \left(7 x^2-2
   x-\frac{2}{1-x}+3\right) \pi^2-24 \left(5 x^2-2
   x-1\right)\right) \lx+8 \left(36 x^3-36 x^2+15
   x-\frac{4}{1-x}+5\right) \li2 \lx
\ibrk
+16 \left(6
   x^3-6 x^2+x\right) \li3 \lx-8 \left(24 x^2-16
   x+\frac{2}{1-x}+1\right) \lm \lx+16 \left(6
   x^3-6 x^2+x\right) \li2 \lm
   \lx
\ibrk
-\frac{4}{3} \left(6 x^3-6 x^2+x\right)
   \lx^3 \ly+16 \left(3 x^2-2 x\right)
   \lm^2 \ly+4 \left(36 x^3-42 x^2+17
   x-\frac{3}{1-x}+3\right) \lx^2 \ly
\ibrk
+8 \left(6
   x^3-6 x^2+x\right) \lm \lx^2
   \ly+\left(120 x^2-\frac{272 x}{3}-\frac{4}{9} \left(126
   x^2-36 x+\frac{22}{1-x}+14-\frac{45}{x}+\frac{27}{x^2}\right)
   \pi^2
\dibrk
+\left(80-\frac{64}{1-x}\right) \z3+\frac{694}{9
   (1-x)}-\frac{478}{9}-\frac{24}{x}\right) \ly+\frac{16
   \li3 \ly}{1-x}+\frac{8}{3} \left(72 x^2-55
   x-\frac{11}{1-x}+37-\frac{9}{x}\right) \lm
   \ly
\ibrk
+\frac{44}{3} \left(1-\frac{1}{1-x}\right)
   \ls \ly-8 \left(6 x^3-6 x^2+x\right)
   \lm^2 \lx \ly+\left(-\frac{4}{3}
   \left(-42 x^3+42 x^2-7 x-\frac{2}{1-x}+4\right) \pi^2
\dibrk
-8
   \left(39 x^3-39 x^2+8 x-\frac{2}{1-x}+1\right)\right) \lx
   \ly-\frac{16 \li2 \lx
   \ly}{1-x}-8 \left(36 x^3-36 x^2+11 x-1\right)
   \lm \lx \ly
\ibrk
-8 \left(-36 x^3+36
   x^2-7 x-\frac{2}{1-x}+3+\frac{5}{x}-\frac{3}{x^2}\right)
   \s12+16 \left(6 x^3-6 x^2+x\right) \lm
   \s12-\frac{16 \lx \s12}{1-x}
\ibrk
-16
   \left(2-\frac{1}{1-x}\right) \ly \s12-16
   \left(-6 x^3+6 x^2-x-\frac{3}{1-x}+5\right) \s13+\frac{16
   \s22}{1-x}
\right]
\right \}
\nonumber \\ &&
+ \cf^2  \left \{
\frac{1}{\text{m_s}^2}
\left[
-\frac{11}{45} (1-x) x \pi^4+\frac{29}{6} (1-x) x
   \pi^2+\frac{255}{8} (1-x) x-30 (1-x) x \z3
\right]
\brk
+
\frac{1}{\text{m_s}}
\left[
-\frac{11}{15} \pi^4+\frac{29 \pi^2}{2}-90
   \z3+\frac{765}{8}
\right]
+
\left[
-\frac{2}{45} \left(-1026 x^3+1026 x^2-171 x+\frac{44}{1-x}\right)
   \pi^4
\ibrk
-\frac{2}{3} \left(468 x^3-468 x^2+54
   x-\frac{57}{1-x}+59-\frac{24}{x}\right) \pi^2+16
   \li2 \pi^2-\frac{2}{3} \left(6 x^3-6
   x^2+x\right) \lx^4
\ibrk
-\frac{2}{3} \left(6 x^3-6
   x^2+x-\frac{3}{1-x}+1\right) \ly^4+\frac{8}{3} \left(36
   x^3-30 x^2+11 x-\frac{3}{1-x}+3\right) \lx^3+\frac{16}{3}
   \left(6 x^3-6 x^2+x\right) \lm
   \lx^3
\ibrk
-\frac{4}{3} \left(-72 x^3+84 x^2-24
   x-\frac{3}{1-x}+11-\frac{12}{x}+\frac{4}{x^2}\right)
   \ly^3+\frac{16}{3} \left(6 x^3-6 x^2+x\right)
   \lm \ly^3
\ibrk
-\frac{8}{3} \left(6 x^3-6
   x^2+x+2\right) \lx \ly^3+\left(-8 \left(6 x^3-6
   x^2+x\right) \pi^2-4 \left(12
   x-\frac{1}{1-x}+1\right)\right) \lm^2
\ibrk
-8 \left(6 x^3-6
   x^2+x\right) \lm^2 \lx^2+\left(\frac{28}{3}
   \left(6 x^3-6 x^2+x\right) \pi^2-8 \left(39 x^3-15
   x^2+x-\frac{4}{1-x}+7\right)\right) \lx^2
\ibrk
+16 \left(6 x^3-6
   x^2+x\right) \li2 \lx^2-8 \left(36 x^3-24 x^2+7
   x-\frac{3}{1-x}+3\right) \lm \lx^2-8 \left(6
   x^3-6 x^2+x\right) \lm^2 \ly^2
\ibrk
+4 \left(6 x^3-6
   x^2+x+\frac{2}{1-x}\right) \lx^2
   \ly^2+\left(\frac{4}{3} \left(42 x^3-42 x^2+7
   x-\frac{6}{1-x}+10\right) \pi^2
\dibrk
+2 \left(-156 x^3+252 x^2-88
   x-\frac{19}{1-x}+17-\frac{16}{x}+\frac{16}{x^2}\right)\right)
   \ly^2
\ibrk
+8 \left(-36 x^3+48 x^2-16
   x+\frac{1}{1-x}+3-\frac{6}{x}+\frac{2}{x^2}\right) \lm
   \ly^2-16 \left(3 x^2-x\right) \lx
   \ly^2
\ibrk
+\frac{1}{2} \left(136 x+\left(1152 x^3-1152 x^2+224
   x-\frac{192}{1-x}+96\right) \z3+\frac{331}{1-x}-331\right)
\ibrk
+16
   \left(36 x^3-36 x^2+15 x-\frac{4}{1-x}+5\right) \li3+32
   \left(6 x^3-6 x^2+x\right) \li4
\ibrk
+\left(-8 \left(36 x^3-36
   x^2+11 x-1\right) \pi^2-36 x+\left(192 x^3-192 x^2+32
   x\right) \z3-\frac{52}{1-x}+192\right) \lm
\ibrk
+32 \left(6
   x^3-6 x^2+x\right) \li3 \lm+32 \left(3 x^2-2
   x\right) \lm^2 \lx
\ibrk
+\left(48 \left(5 x^2-2
   x-1\right)-16 \left(7 x^2-2 x-\frac{2}{1-x}+3\right)
   \pi^2\right) \lx-16 \left(36 x^3-36 x^2+15
   x-\frac{4}{1-x}+5\right) \li2 \lx
\ibrk
-32 \left(6
   x^3-6 x^2+x\right) \li3 \lx+16 \left(24 x^2-16
   x+\frac{2}{1-x}+1\right) \lm \lx-32 \left(6
   x^3-6 x^2+x\right) \li2 \lm
   \lx
\ibrk
+\frac{8}{3} \left(6 x^3-6 x^2+x\right)
   \lx^3 \ly-32 \left(3 x^2-2 x\right)
   \lm^2 \ly-8 \left(36 x^3-42 x^2+17
   x-\frac{3}{1-x}+3\right) \lx^2 \ly
\ibrk
-16 \left(6
   x^3-6 x^2+x\right) \lm \lx^2
   \ly+\left(-240 x^2+232 x+\frac{8}{3} \left(42 x^2-13
   x-\frac{6}{1-x}+13-\frac{18}{x}+\frac{6}{x^2}\right)
   \pi^2
\dibrk
+\left(\frac{80}{1-x}-112\right)
   \z3-\frac{50}{1-x}+10+\frac{32}{x}\right) \ly-\frac{32
   \li3 \ly}{1-x}-8 \left(48 x^2-41
   x-\frac{2}{1-x}+15-\frac{4}{x}\right) \lm
   \ly
\ibrk
+16 \left(6 x^3-6 x^2+x\right) \lm^2
   \lx \ly+\left(\frac{8}{3} \left(-42 x^3+42
   x^2-7 x-\frac{2}{1-x}+4\right) \pi^2
\dibrk
+16 \left(39 x^3-39
   x^2+8 x-\frac{2}{1-x}+1\right)\right) \lx
   \ly+\frac{32 \li2 \lx
   \ly}{1-x}+16 \left(36 x^3-36 x^2+11 x-1\right)
   \lm \lx \ly
\ibrk
-8 \left(72 x^3-72
   x^2+16 x-\frac{1}{1-x}+1-\frac{12}{x}+\frac{4}{x^2}\right)
   \s12-32 \left(6 x^3-6 x^2+x\right) \lm
   \s12+\frac{32 \lx \s12}{1-x}
\ibrk
+16
   \left(3-\frac{1}{1-x}\right) \ly \s12+16
   \left(-12 x^3+12 x^2-2 x-\frac{3}{1-x}+7\right) \s13+32
   \s22
\right]
 \right \} 
\nonumber \\ &&
+ \nf \tf \cf \left \{
\frac{1}{\text{m_s}^2}
\left[
\frac{7}{9} (1-x) x \pi^2+\frac{4085}{162} (1-x) x-\frac{2}{9}
   (1-x) x \z3-\frac{32}{3} (1-x) x \ls
\right]
\brk
+
\frac{1}{\text{m_s}}
\left[
\frac{7 \pi^2}{3}-\frac{2 \z3}{3}-32
   \ls+\frac{4085}{54}
\right]
+
\left[
-\frac{44}{45} \left(1-\frac{2}{1-x}\right) \pi^4-\frac{4}{9}
   \left(80 x+\frac{12}{1-x}+22-\frac{27}{x}\right)
   \pi^2
\ibrk
+\frac{16}{3} \left(1-\frac{2}{1-x}\right)
   \li2 \pi^2+\frac{4}{3} \left(2 x^2-2
   x-\frac{4}{1-x}+\frac{3}{(x-1)^2}+1\right) \lx^3
\ibrk
-\frac{4}{9}
   \left(6 x^2-6 x-\frac{8}{1-x}+11-\frac{12}{x}+\frac{9}{x^2}\right)
   \ly^3-4 \left(8 x-\frac{10}{1-x}+\frac{3}{(x-1)^2}+4\right)
   \lx^2
\ibrk
-4 \left(\frac{3}{(x-1)^2}+1-\frac{4}{1-x}\right)
   \lm \lx^2+\frac{4}{9} \left(-72
   x-\frac{26}{1-x}+152-\frac{117}{x}+\frac{54}{x^2}\right)
   \ly^2+4 \left(1-\frac{4}{x}+\frac{3}{x^2}\right)
   \lm \ly^2
\ibrk
-\frac{16}{3}
   \left(1-\frac{1}{1-x}\right) \ls
   \ly^2-\frac{4}{3} \left(4 x^2-2 x+1\right) \lx
   \ly^2-\frac{2}{81} \left(-1368 x
\dibrk
+\left(1080 x^2-2160
   x-\frac{720}{1-x}-\frac{972}{(x-1)^2}+2664\right)
   \z3-\frac{4769}{1-x}+4769\right)
\ibrk
+\frac{8}{3} \left(10 x^2-8
   x-\frac{7}{1-x}-\frac{9}{(x-1)^2}+19\right)
   \li3+\left(\frac{16}{9} \left(13 x-\frac{13}{1-x}+32\right)-4
   (2 x-1) \pi^2\right) \lm
\ibrk
+\frac{16}{3} \left(-2
   x-\frac{9}{1-x}+9\right) \ls-\frac{32}{3}
   \left(x-\frac{1}{1-x}+2\right) \lm \ls-16
   \left(x^2-x\right) \lm^2 \lx
\ibrk
+\left(-\frac{4}{9}
   \left(32 x^2-28 x-\frac{41}{1-x}+\frac{27}{(x-1)^2}+32\right)
   \pi^2-\frac{8}{3} \left(44 x^2-18 x-9\right)\right)
   \lx
\ibrk
-\frac{8}{3} \left(10 x^2-8
   x-\frac{7}{1-x}-\frac{9}{(x-1)^2}+19\right) \li2
   \lx-8 \left(16 x^2-14 x+\frac{3}{1-x}\right)
   \lm \lx+16 \left(x^2-x\right)
   \lm^2 \ly
\ibrk
-4 \left(2 x^2-2
   x-\frac{1}{1-x}-\frac{3}{(x-1)^2}+4\right) \lx^2
   \ly+\left(\frac{4}{9} \left(32 x^2-16
   x+\frac{8}{1-x}+12-\frac{36}{x}+\frac{27}{x^2}\right)
   \pi^2
\dibrk
+\frac{8}{9} \left(132 x^2-198
   x-\frac{31}{1-x}+58+\frac{27}{x}\right)\right)
   \ly-\frac{8}{3} \left(-48 x^2+58
   x-\frac{4}{1-x}+2-\frac{9}{x}\right) \lm
   \ly
\ibrk
-\frac{16}{3} \left(1-\frac{1}{1-x}\right)
   \ls \ly+\frac{8}{3} (20 x-3) \lx
   \ly+8 (2 x-1) \lm \lx
   \ly+\frac{8}{3} \left(10 x^2-8
   x-11+\frac{12}{x}-\frac{9}{x^2}\right) \s12
\ibrk
+32
   \left(1-\frac{2}{1-x}\right) \s22
\right]
\right \} \, ,
\eea

\bea
&&{\cal J}_2^{(0 \times 2)} (m_s,x) =  \nonumber \\
&& \ca \cf \left \{
\frac{1}{\text{m_s}^2}
\left[
\frac{31}{120} (1-x) x \pi^4-\frac{107}{36} (1-x) x
   \pi^2-\frac{51157}{648} (1-x) x+\frac{659}{18} (1-x) x
   \z3+\frac{88}{3} (1-x) x \ls
\right]
\brk
+
\frac{1}{\text{m_s}}
\left[
\frac{31}{120} \left(2 x^2-2 x+3\right) \pi^4-\frac{107}{36}
   \left(2 x^2-2 x+3\right) \pi^2-\frac{51157}{648} \left(2
   x^2-2 x+3\right)+\frac{659}{18} \left(2 x^2-2 x+3\right)
   \z3
\ibrk
+\frac{88}{3} \left(2 x^2-2 x+3\right) \ls
\right]
+
\left[
\frac{1}{60} \left(-684 x^3+684 x^2-52 x-93\right)
   \pi^4+\frac{1}{18} \left(1296 x^3-1224 x^2-346
   x
\dibrk
-\frac{88}{1-x}+465-\frac{108}{x}\right) \pi^2+\frac{8}{3}
   \li2 \pi^2+\frac{1}{6} \left(6 x^3-6
   x^2+x\right) \lx^4+\frac{1}{6} \left(6 x^3-6 x^2-x+8\right)
   \ly^4
\ibrk
-\frac{2}{3} \left(36 x^3-30 x^2+11
   x-\frac{3}{1-x}+3\right) \lx^3-\frac{4}{3} \left(6 x^3-6
   x^2+x\right) \lm \lx^3
\ibrk
-\frac{2}{9} \left(108
   x^3-126 x^2+34 x-56+\frac{15}{x}-\frac{9}{x^2}\right)
   \ly^3-\frac{4}{3} \left(6 x^3-6 x^2+x\right)
   \lm \ly^3+\frac{2}{3} \left(6 x^3-6 x^2+3
   x-8\right) \lx \ly^3
\ibrk
+\left(2 \left(6 x^3-6
   x^2+x\right) \pi^2+12 x\right) \lm^2+2 \left(6
   x^3-6 x^2+x\right) \lm^2 \lx^2+\left(2 \left(36
   x^3-10 x^2-6 x+\frac{1}{1-x}+2\right)
\dibrk
-\frac{7}{3} \left(6 x^3-6
   x^2+x\right) \pi^2\right) \lx^2-4 \left(6
   x^3-6 x^2+x\right) \li2 \lx^2+2 \left(36 x^3-24
   x^2+7 x-\frac{3}{1-x}+3\right) \lm \lx^2
\ibrk
+2
   \left(6 x^3-6 x^2+x\right) \lm^2 \ly^2 +\left(-6
   x^3+6 x^2-5 x+12\right) \lx^2
   \ly^2+\left(\frac{1}{3} \left(-42 x^3+42 x^2-13 x+20\right)
   \pi^2
\dibrk
+\frac{1}{9} \left(648 x^3-1044 x^2+491
   x+\frac{132}{1-x}-634+\frac{198}{x}-\frac{108}{x^2}\right)\right)
   \ly^2+2 \left(36 x^3-48 x^2+15
   x-5+\frac{5}{x}-\frac{3}{x^2}\right) \lm
   \ly^2
\ibrk
-\frac{22}{3} (x-2) \ls
   \ly^2+6 \left(2 x^2-3 x+1\right) \lx
   \ly^2+\frac{1}{324} \left(-110054 x+\left(-46656 x^3+46656
   x^2+28008 x+\frac{6480}{1-x}
\ddibrk
-73764\right)
   \z3-\frac{15480}{1-x}+168951\right)-4 \left(36 x^3-36 x^2+24
   x+\frac{1}{1-x}\right) \li3-8 \left(6 x^3-6 x^2+x\right)
   \li4
\ibrk
+\left(2 \left(36 x^3-36 x^2+11 x-1\right)
   \pi^2-\frac{82 x}{9}+\left(-48 x^3+48 x^2-8 x\right)
   \z3+\frac{298}{9 (1-x)}-\frac{944}{9}\right) \lm
\ibrk
-8
   \left(6 x^3-6 x^2+x\right) \li3
   \lm-\frac{22}{3} \left(-17 x-\frac{2}{1-x}+26\right)
   \ls+\frac{44}{3} \left(x-\frac{1}{1-x}+2\right)
   \lm \ls
\ibrk
-8 \left(3 x^2-2 x\right)
   \lm^2 \lx+\left(\frac{2}{3} \left(42 x^2-7
   x-\frac{11}{1-x}+17\right) \pi^2+12 \left(-4 x^2+2
   x-\frac{1}{1-x}+2\right)\right) \lx
\ibrk
+4 \left(36 x^3-36 x^2+24
   x+\frac{1}{1-x}\right) \li2 \lx+8 \left(6 x^3-6
   x^2+x\right) \li3 \lx-4 \left(24 x^2-16
   x+\frac{2}{1-x}+1\right) \lm \lx
\ibrk
+8 \left(6
   x^3-6 x^2+x\right) \li2 \lm
   \lx-\frac{2}{3} \left(6 x^3-6 x^2+x\right)
   \lx^3 \ly+8 \left(3 x^2-2 x\right)
   \lm^2 \ly
\ibrk
-4 \left(-18 x^3+21 x^2-13
   x-\frac{1}{1-x}+1\right) \lx^2 \ly+4 \left(6
   x^3-6 x^2+x\right) \lm \lx^2
   \ly
\ibrk
+\left(48 x^2-\frac{433 x}{9}-\frac{2}{9} \left(126 x^2-62
   x+28-\frac{45}{x}+\frac{27}{x^2}\right) \pi^2+(32-24 x)
   \z3-\frac{166}{9 (1-x)}+\frac{382}{9}-\frac{12}{x}\right)
   \ly
\ibrk
+16 (x-3) \li3 \ly+\frac{4}{3}
   \left(72 x^2-55 x-\frac{11}{1-x}+37-\frac{9}{x}\right) \lm
   \ly-\frac{22}{3} \left(x-\frac{2}{1-x}+2\right)
   \ls \ly
\ibrk
-4 \left(6 x^3-6 x^2+x\right)
   \lm^2 \lx \ly+\left(\frac{2}{3}
   \left(42 x^3-42 x^2+7 x+4\right) \pi^2-4 \left(36 x^3-34
   x^2+2 x-\frac{2}{1-x}+1\right)\right) \lx
   \ly
\ibrk
-16 (x-3) \li2 \lx
   \ly-4 \left(36 x^3-36 x^2+11 x-1\right) \lm
   \lx \ly-4 \left(-36 x^3+36 x^2-12
   x+\frac{1}{1-x}
\dibrk
+\frac{5}{x}-\frac{3}{x^2}\right) \s12+8
   \left(6 x^3-6 x^2+x\right) \lm \s12-16 (x-3)
   \lx \s12+16 \ly
   \s12
\ibrk
+16 \left(3 x^3-3 x^2+x+1\right) \s13 +16
   (x-3) \s22
\right]
\right \}
\nonumber \\ &&
+ \cf^2  \left \{
\frac{1}{\text{m_s}^2}
\left[
-\frac{11}{45} (1-x) x \pi^4+\frac{29}{6} (1-x) x
   \pi^2+\frac{255}{8} (1-x) x-30 (1-x) x \z3
\right]
\brk
+
\frac{1}{\text{m_s}}
\left[
-\frac{11}{45} \left(2 x^2-2 x+3\right) \pi^4+\frac{29}{6}
   \left(2 x^2-2 x+3\right) \pi^2+\frac{255}{8} \left(2 x^2-2
   x+3\right)-30 \left(2 x^2-2 x+3\right) \z3
\right]
\brk
+
\left[
\frac{1}{45} \left(1026 x^3-1026 x^2+149 x+22\right)
   \pi^4+\frac{1}{3} \left(-432 x^3+408 x^2+76
   x+\frac{72}{1-x}-161+\frac{24}{x}\right) \pi^2-\frac{8}{3} x
   \li2 \pi^2
\ibrk
+\frac{1}{3} \left(-6 x^3+6
   x^2-x\right) \lx^4-\frac{2}{3} \left(3 x^3-3 x^2-2 x+7\right)
   \ly^4+\frac{4}{3} \left(36 x^3-30 x^2+11
   x-\frac{3}{1-x}+3\right) \lx^3
\ibrk
+\frac{8}{3} \left(6 x^3-6
   x^2+x\right) \lm \lx^3-\frac{2}{3} \left(-72
   x^3+84 x^2-11 x+\frac{6}{1-x}+2-\frac{12}{x}+\frac{4}{x^2}\right)
   \ly^3+\frac{8}{3} \left(6 x^3-6 x^2+x\right)
   \lm \ly^3
\ibrk
-\frac{4}{3} \left(6 x^3-6 x^2+3
   x-8\right) \lx \ly^3+\left(-4 \left(6 x^3-6
   x^2+x\right) \pi^2-2 \left(12
   x-\frac{1}{1-x}+1\right)\right) \lm^2
\ibrk
-4 \left(6 x^3-6
   x^2+x\right) \lm^2 \lx^2+\left(\frac{14}{3}
   \left(6 x^3-6 x^2+x\right) \pi^2-4 \left(36 x^3-10 x^2-6
   x+\frac{1}{1-x}+2\right)\right) \lx^2
\ibrk
+8 \left(6 x^3-6
   x^2+x\right) \li2 \lx^2-4 \left(36 x^3-24 x^2+7
   x-\frac{3}{1-x}+3\right) \lm \lx^2-4 \left(6
   x^3-6 x^2+x\right) \lm^2 \ly^2
\ibrk
+2 \left(6 x^3-6
   x^2+5 x-12\right) \lx^2 \ly^2+\left(-144
   x^3+232 x^2-83 x+\frac{2}{3} \left(42 x^3-42 x^2+5 x-4\right)
   \pi^2-\frac{14}{1-x}
\dibrk
+46-\frac{24}{x}+\frac{16}{x^2}\right)
   \ly^2+4 \left(-36 x^3+48 x^2-16
   x+\frac{1}{1-x}+3-\frac{6}{x}+\frac{2}{x^2}\right) \lm
   \ly^2-12 \left(2 x^2-3 x+1\right) \lx
   \ly^2
\ibrk
+\frac{1}{4} \left(586 x+\left(1152 x^3-1152 x^2-48
   x-\frac{64}{1-x}+688\right) \z3+\frac{144}{1-x}-909\right)
\ibrk
+8 \left(36
   x^3-36 x^2+24 x+\frac{1}{1-x}\right) \li3+16 \left(6 x^3-6
   x^2+x\right) \li4+\left(-4 \left(36 x^3-36 x^2+11 x-1\right)
   \pi^2
\dibrk
-18 x+\left(96 x^3-96 x^2+16 x\right)
   \z3-\frac{26}{1-x}+96\right) \lm+16 \left(6 x^3-6
   x^2+x\right) \li3 \lm+16 \left(3 x^2-2 x\right)
   \lm^2 \lx
\ibrk
+\left(-\frac{4}{3} \left(42 x^2-7
   x-\frac{11}{1-x}+17\right) \pi^2-24 \left(-4 x^2+2
   x-\frac{1}{1-x}+2\right)\right) \lx
\ibrk
-8 \left(36 x^3-36 x^2+24
   x+\frac{1}{1-x}\right) \li2 \lx-16 \left(6
   x^3-6 x^2+x\right) \li3 \lx
\ibrk
+8 \left(24 x^2-16
   x+\frac{2}{1-x}+1\right) \lm \lx-16 \left(6
   x^3-6 x^2+x\right) \li2 \lm
   \lx+\frac{4}{3} \left(6 x^3-6 x^2+x\right)
   \lx^3 \ly
\ibrk
-16 \left(3 x^2-2 x\right)
   \lm^2 \ly+8 \left(-18 x^3+21 x^2-13
   x-\frac{1}{1-x}+1\right) \lx^2 \ly-8 \left(6
   x^3-6 x^2+x\right) \lm \lx^2
   \ly
\ibrk
+\left(-96 x^2+127 x+\frac{8}{3} \left(21 x^2-16
   x+\frac{3}{1-x}+11-\frac{9}{x}+\frac{3}{x^2}\right)
   \pi^2+(24 x-16)
   \z3+\frac{14}{1-x}-50+\frac{16}{x}\right) \ly
\ibrk
-32 (x-3)
   \li3 \ly-4 \left(48 x^2-41
   x-\frac{2}{1-x}+15-\frac{4}{x}\right) \lm \ly+8
   \left(6 x^3-6 x^2+x\right) \lm^2 \lx
   \ly
\ibrk
+\left(8 \left(36 x^3-34 x^2+2
   x-\frac{2}{1-x}+1\right)-\frac{4}{3} \left(42 x^3-42 x^2+7 x+4\right)
   \pi^2\right) \lx \ly+32 (x-3)
   \li2 \lx \ly
\ibrk
+8 \left(36 x^3-36
   x^2+11 x-1\right) \lm \lx \ly-4
   \left(72 x^3-72 x^2+23 x-\frac{4}{1-x}+10-\frac{12}{x}+\frac{4}{x^2}\right)
   \s12
\ibrk
-16 \left(6 x^3-6 x^2+x\right) \lm
   \s12+32 (x-3) \lx \s12+8 (x-6)
   \ly \s12-8 \left(12 x^3-12 x^2+x+10\right)
   \s13
\ibrk
-16 (3 x-8) \s22
\right]
\right \} 
\nonumber \\ &&
+ \nf \tf \cf \left \{
\frac{1}{\text{m_s}^2}
\left[
\frac{7}{9} (1-x) x \pi^2+\frac{4085}{162} (1-x) x-\frac{2}{9}
   (1-x) x \z3-\frac{32}{3} (1-x) x \ls
\right] 
\brk
+
\frac{1}{\text{m_s}}
\left[
\frac{7}{9} \left(2 x^2-2 x+3\right) \pi^2+\frac{4085}{162}
   \left(2 x^2-2 x+3\right)-\frac{2}{9} \left(2 x^2-2 x+3\right)
   \z3-\frac{32}{3} \left(2 x^2-2 x+3\right) \ls
\right]
\brk
+
\left[
\frac{44}{45} (x+1) \pi^4-\frac{2}{9} \left(-26
   x+\frac{58}{1-x}+27-\frac{54}{x}\right) \pi^2-16
   \li2 \pi^2+\frac{4}{3} \left(2 x^2-2
   x-\frac{4}{1-x}+\frac{3}{(x-1)^2}+1\right) \lx^3
\ibrk
-\frac{4}{9}
   \left(6 x^2-10 x+11-\frac{12}{x}+\frac{9}{x^2}\right) \ly^3-4
   \left(10 x-\frac{16}{1-x}+\frac{6}{(x-1)^2}+7\right) \lx^2-4
   \left(\frac{3}{(x-1)^2}+1-\frac{4}{1-x}\right) \lm
   \lx^2
\ibrk
-8 (x-2) \lx^2
   \ly^2+\frac{4}{9} \left(-103
   x-\frac{12}{1-x}+203-\frac{144}{x}+\frac{54}{x^2}\right)
   \ly^2+4 \left(1-\frac{4}{x}+\frac{3}{x^2}\right)
   \lm \ly^2+\frac{8}{3} (x-2) \ls
   \ly^2
\ibrk
-4 (2 x-1) \lx
   \ly^2+\frac{1}{81} \left(8854 x+\left(-1296 x^2+10296
   x-\frac{2592}{1-x}+\frac{1944}{(x-1)^2}-1188\right)
   \z3+\frac{1368}{1-x}-13623\right)
\ibrk
+8 \left(2 x^2-12
   x+\frac{4}{1-x}-\frac{3}{(x-1)^2}\right)
   \li3+\left(\frac{8}{9} \left(13 x-\frac{13}{1-x}+32\right)-4
   (2 x-1) \pi^2\right) \lm
\ibrk
+\frac{8}{3} \left(-17
   x-\frac{2}{1-x}+26\right) \ls-\frac{16}{3}
   \left(x-\frac{1}{1-x}+2\right) \lm \ls-16
   \left(x^2-x\right) \lm^2 \lx+\left(24 \left(-4
   x^2+2 x-\frac{1}{1-x}+2\right)
\dibrk
-\frac{4}{3} \left(4 x^2-8
   x-\frac{12}{1-x}+\frac{9}{(x-1)^2}+9\right) \pi^2\right)
   \lx-8 \left(2 x^2-12 x+\frac{4}{1-x}-\frac{3}{(x-1)^2}\right)
   \li2 \lx
\ibrk
-8 \left(16 x^2-14
   x+\frac{3}{1-x}\right) \lm \lx+16
   \left(x^2-x\right) \lm^2 \ly+4 \left(-2 x^2+10
   x-\frac{4}{1-x}+\frac{3}{(x-1)^2}+1\right) \lx^2
   \ly
\ibrk
+\left(96 x^2-\frac{1372 x}{9}+\frac{4}{9} \left(12 x^2+4
   x+7-\frac{36}{x}+\frac{27}{x^2}\right) \pi^2+(64-32 x)
   \z3+\frac{56}{9 (1-x)}-\frac{56}{9}+\frac{24}{x}\right)
   \ly
\ibrk
+32 (x-2) \li3 \ly+\frac{8}{3}
   \left(48 x^2-56 x+\frac{2}{1-x}+2+\frac{9}{x}\right) \lm
   \ly+\frac{8}{3} \left(x-\frac{2}{1-x}+2\right)
   \ls \ly
\ibrk
+\left(\frac{16}{3} (x-2)
   \pi^2+8 (2 x-1)\right) \lx
   \ly-32 (x-2) \li2 \lx
   \ly+8 (2 x-1) \lm \lx
   \ly
\ibrk
+8 \left(2 x^2+8 x-10+\frac{4}{x}-\frac{3}{x^2}\right)
   \s12-32 (x-2) \lx \s12-96
   \s22
\right]
\right \} \, ,
\eea

\bea
&&{\cal J}_3^{(0 \times 2)} (m_s,x) =  \nonumber \\
&& \ca \cf \left \{
\frac{1}{\text{m_s}^2}
\left[
\frac{31}{240} (1-x) x \pi^4-\frac{107}{72} (1-x) x
   \pi^2-\frac{51157 (1-x) x}{1296}+\frac{659}{36} (1-x) x
   \z3+\frac{44}{3} (1-x) x \ls
\right]
\brk
+
\frac{1}{\text{m_s}}
\left[
\frac{31}{240} \left(4 x^2-4 x+3\right) \pi^4-\frac{107}{72}
   \left(4 x^2-4 x+3\right) \pi^2-\frac{51157 \left(4 x^2-4
   x+3\right)}{1296}+\frac{659}{36} \left(4 x^2-4 x+3\right)
   \z3
\ibrk
+\frac{44}{3} \left(4 x^2-4 x+3\right) \ls
\right]
+
\left[
-\frac{31}{20} \left(x^2-x+1\right) \pi^4+\frac{107}{6}
   \left(x^2-x+1\right) \pi^2+\frac{1}{108} \left(51157
   x^2-51157 x
\dibrk
+\left(-23724 x^2+23724 x-23724\right)
   \z3+51157\right)-176 \left(x^2-x+1\right) \ls
\right]
\right \}
\nonumber \\ &&
+ \cf^2  \left \{
\frac{1}{\text{m_s}^2}
\left[
-\frac{11}{90} (1-x) x \pi^4+\frac{29}{12} (1-x) x
   \pi^2+\frac{255}{16} (1-x) x-15 (1-x) x \z3
\right]
\brk
+
\frac{1}{\text{m_s}}
\left[
-\frac{11}{90} \left(4 x^2-4 x+3\right) \pi^4+\frac{29}{12}
   \left(4 x^2-4 x+3\right) \pi^2+\frac{255}{16} \left(4 x^2-4
   x+3\right)-15 \left(4 x^2-4 x+3\right) \z3
\right]
\brk
+
\left[
\frac{22}{15} \left(x^2-x+1\right) \pi^4-29
   \left(x^2-x+1\right) \pi^2+\frac{45}{4} \left(-17 x^2+17
   x+\left(16 x^2-16 x+16\right) \z3-17\right)
\right]
\right \} 
\nonumber \\ &&
+ \nf \tf \cf \left \{
\frac{1}{\text{m_s}^2}
\left[
\frac{7}{18} (1-x) x \pi^2+\frac{4085}{324} (1-x) x-\frac{1}{9}
   (1-x) x \z3-\frac{16}{3} (1-x) x \ls
\right]
\brk
+
\frac{1}{\text{m_s}}
\left[
\frac{7}{18} \left(4 x^2-4 x+3\right) \pi^2+\frac{4085}{324}
   \left(4 x^2-4 x+3\right)+\frac{1}{9} \left(-4 x^2+4 x-3\right)
   \z3-\frac{16}{3} \left(4 x^2-4 x+3\right) \ls
\right]
\brk
+
\left[
-\frac{14}{3} \left(x^2-x+1\right) \pi^2+\frac{1}{27}
   \left(-4085 x^2+4085 x+\left(36 x^2-36 x+36\right) \z3-4085\right)+64
   \left(x^2-x+1\right) \ls
\right]
\right \} \, , \nonumber \\
\eea

\bea
\label{eq:theendend}
&&{\cal J}_4^{(0 \times 2)} (m_s,x) =  \nonumber \\
&& \nf \tf \cf \left \{
\left[
\frac{22}{45} (x+1) \pi^4-\frac{2}{3} \left(-2
   x+\frac{9}{1-x}+1-\frac{9}{x}\right) \pi^2-8
   \li2 \pi^2-2 (2 x-1) \lm
   \pi^2
\ibrk
+\frac{2}{3} \left(2 x^2-2
   x-\frac{4}{1-x}+\frac{3}{(x-1)^2}+1\right) \lx^3-\frac{2}{3}
   \left(2 x^2-2 x+1-\frac{4}{x}+\frac{3}{x^2}\right) \ly^3
\ibrk
-2
   \left(10 x-\frac{16}{1-x}+\frac{6}{(x-1)^2}+7\right) \lx^2-2
   \left(\frac{3}{(x-1)^2}+1-\frac{4}{1-x}\right) \lm
   \lx^2-4 (x-2) \lx^2 \ly^2
\ibrk
+2
   \left(-10 x+17-\frac{16}{x}+\frac{6}{x^2}\right) \ly^2+2
   \left(1-\frac{4}{x}+\frac{3}{x^2}\right) \lm
   \ly^2-2 (2 x-1) \lx \ly^2
\ibrk
-4
   \left(2 x^2-16 x+\frac{4}{1-x}-\frac{3}{(x-1)^2}+2\right) \z3+4
   \left(2 x^2-12 x+\frac{4}{1-x}-\frac{3}{(x-1)^2}\right)
   \li3
\ibrk
-8 \left(x^2-x\right) \lm^2
   \lx+\left(12 \left(-4 x^2+2
   x-\frac{1}{1-x}+2\right)-\frac{2}{3} \left(4 x^2-8
   x-\frac{12}{1-x}+\frac{9}{(x-1)^2}+9\right) \pi^2\right)
   \lx
\ibrk
-4 \left(2 x^2-12 x+\frac{4}{1-x}-\frac{3}{(x-1)^2}\right)
   \li2 \lx-4 \left(16 x^2-14
   x+\frac{3}{1-x}\right) \lm \lx+8
   \left(x^2-x\right) \lm^2 \ly
\ibrk
+2 \left(-2 x^2+10
   x-\frac{4}{1-x}+\frac{3}{(x-1)^2}+1\right) \lx^2
   \ly+\left(48 x^2-72 x+\frac{2}{3} \left(4
   x^2+5-\frac{12}{x}+\frac{9}{x^2}\right) \pi^2
\dibrk
+(32-16 x)
   \z3+\frac{12}{x}\right) \ly+16 (x-2) \li3
   \ly+4 \left(16 x^2-18 x+2+\frac{3}{x}\right)
   \lm \ly
\ibrk
+\left(\frac{8}{3} (x-2)
   \pi^2+4 (2 x-1)\right) \lx
   \ly-16 (x-2) \li2 \lx
   \ly+4 (2 x-1) \lm \lx
   \ly
\ibrk
+4 \left(2 x^2+8 x-10+\frac{4}{x}-\frac{3}{x^2}\right)
   \s12-16 (x-2) \lx \s12-48
   \s22
\right]
\right \} \, ,
\eea
\normalsize

where $\lm$, $\ls$, $\lx$ and $\ly$ are defined as
\begin{equation}
\lm = \log\left( m_s \right)\, , \;\;\;\;
\ls = \log\left( \frac{s}{\mu^2} \right)\, , \;\;\;\;
\lx = \log\left( x \right)\, , \;\;\;\;
\ly = \log\left( 1-x \right)\, .
\end{equation}

\subsection{One-loop Squared Contribution}
In this section, we give explicit expressions for the finite remainder of 
the one-loop squared contribution ${\cal F}_{\!inite,}^{( 1 \times 1)}$
defined as 
\begin{equation}
 {\cal F}_{\!inite,}^{( 1 \times 1)}(s, t, u, m, \mu) = \A^{\nnlo \, 
(1 \times 1)}(s, t, u, m, \mu) 
     - \mathcal{C}_{atani}^{(1 \times 1)}(s, t, u, m, \mu)\, ,
\end{equation} 

The EW structure of the finite remainder for the one-loop squared
corrections, similarly to the case of the two-loop corrections,
can be factorised as 
\begin{eqnarray}
\label{eq:down}
{\cal F}_{\!inite, \,\, down}^{\, (1 \times 1)} &=& 
N \, \cf^2
\sum_{i=1, 3}\, c^{}_{i}  
{\cal J}_{i,\,down}^{(1 \times 1)}(m_s, x, \frac{s}{\mu^2})  \, .
\end{eqnarray}

Our result then reads:
\footnotesize
\bea
{\mathcal J_1}^{(1 \times 1)} &=&
\frac{64 (1-x) x}{\text{m_s}^2}
+
\frac{192}{\text{m_s}}
+
\left[
-4 \left(-\frac{2}{x}-\frac{1}{x^2}-\frac{1}{x^3}+1-\frac{1}{1-x}\right)
   \ly^4+8 \left(\frac{3}{x}+\frac{2}{x^2}+\frac{3}{1-x}\right)
   \ly^3
\brk
+\left(4 \left(\frac{6}{x}+15-\frac{5}{1-x}\right)-16
   \left(-\frac{2}{x}-\frac{1}{x^2}-\frac{1}{x^3}+1-\frac{1}{1-x}\right)
   \pi^2\right) \ly^2
\brk
+\left(16
   \left(\frac{3}{x}+\frac{2}{x^2}+\frac{3}{1-x}\right) \pi^2+8
   \left(7-\frac{5}{1-x}\right)\right) \ly-4
   \left(-\frac{4}{x}+1-\frac{9}{1-x}\right) \pi^2
\brk
-4 \left(-32
   x-\frac{83}{1-x}+83\right)+128 \left(x-\frac{1}{1-x}+2\right)
   \lm
\right]
\, ,\nonumber \\
{\mathcal J_2}^{(1 \times 1)} &=& 
\frac{64 (1-x) x}{\text{m_s}^2}
+
\frac{1}{\text{m_s}}
\left[
64 \left(2 x^2-2 x+3\right)
\right]
+
\left[
-32 (x-2) \ly^2-32 \left(x-\frac{2}{1-x}+2\right)
   \ly
\brk
-32 \left(-9 x-\frac{2}{1-x}+14\right)+64
   \left(x-\frac{1}{1-x}+2\right) \lm
\right]
\, , \nonumber \\
{\mathcal J_3}^{(1 \times 1)} &=&
\frac{32 (1-x) x}{\text{m_s}^2}
+
\frac{1}{\text{m_s}}
\left[
32 \left(4 x^2-4 x+3\right)
\right]
+
\left[
-384 \left(x^2-x+1\right)
\right] \, .
\eea
\normalsize

%
%
\section{Conclusions}
\label{sec:conclusions}

In this work we have calculated the NNLO QCD virtual corrections for
the process $q {\bar q} \rightarrow W^+ \, W^-$ in the limit of small vector
boson mass. The $\MSbar$  renormalised amplitude is
still infrared divergent and contains poles up to $\mathcal{O}(1/\ep^4)$. 
We checked that the infrared structure of our result
agrees with the prediction of Catani's
formalism for the  infrared structure of QCD amplitudes.  

The main result of our paper has been given as the 
finite remainder of the NNLO 
two-loop and one-loop 
virtual corrections after subtraction of the structure predicted 
by Catani's formalism. This is a first step towards the
complete evaluation of the
virtual corrections. In a forthcoming publication, we will derive a series
expansion in the mass and integrate the result numerically. This will
require the present result as a starting point.

To complete the NNLO project one still needs to consider 
$2 \to 3$ real-virtual contributions  
and $2 \to 4$ real ones. The real-virtual corrections
are known from the NLO studies on $W W + jet$ production
in Refs.~\cite{Campbell:2007ev,
Dittmaier:2007th}. The integration over the full phase space
would require additional subtraction terms, similar to those constructed
in Ref.~\cite{Catani:2007vq}.

%
%
{\bf{Acknowledgments:}}

    This work was supported by the Sofja Kovalevskaja Award of the
    Alexander von Humboldt Foundation and by the German Federal
    Ministry of Education and Research (BMBF) under contract number 05HT6WWA.

%
%
%
%
\setcounter{section}{0}
\renewcommand\thesection{Appendix:}

\section{$\langle {\cal
  M}^{(0)} | {\cal M}^{(1)}_{\rm finite} \rangle $ to order $\epsilon^2$}
\label{app:oneloopep2}

Here we present the expression for the one-loop result,
$\langle {\cal  M}^{(0)} | {\cal M}^{(1)}_{\rm finite} \rangle$ 
up to order $\epsilon^2$
for down-type quarks. 
This result completes the list of the elements needed
in Eq.~(\ref{eq:Aexp}) in order to have the perturbative expansion of the 
amplitude up to order $\alpha_s^2$ in the high energy limit.

\bea
\langle {\cal  M}^{(0)} | {\cal M}^{(1)}_{\rm
  finite} \rangle &=& N C_F \sum_{i= 1,3} c_i {\mathcal J^{(0 \times 1)}_i} \, .
\eea

{\scalefont{0.9}

\bea
{\mathcal J^{(0 \times 1)}_1} &=& 
\left \{
\frac{1}{m_s^2}
\left[
-16 (1-x) x
\right]
+
\frac{1}{m_s}
\left[
-48
\right]
+
\left[
-8 \left(1-\frac{1}{1-x}\right) \ly^2-8
   \left(1-\frac{1}{1-x}\right) \ly
\ibrk
+8 \left(-2
   x-\frac{9}{1-x}+9\right)-16 \left(x-\frac{1}{1-x}+2\right)
   \lm
\right]
\right \} \nonumber \\ &&
+ i \pi \left \{
-16 \ly \left(1-\frac{1}{1-x}\right)-8
   \left(1-\frac{1}{1-x}\right)
\right \} \nonumber \\ &&
+\epsilon 
\left \{
\frac{1}{m_s^2}
\left[
-32 (1-x) x+8 (1-x) \z3 x+16 (1-x) \ls x
\right]
+
\frac{1}{m_s}
\left[
24 \z3+48 \ls-32
\right]
\brk
+
\left[
\frac{16}{3} \left(1-\frac{1}{1-x}\right) \ly^3+4
   \left(3-\frac{5}{1-x}\right) \ly^2+8
   \left(1-\frac{1}{1-x}\right) \ls \ly^2+
\ibrk
\left(8
   \left(1-\frac{1}{1-x}\right) \pi^2
-8
   \left(1+\frac{1}{1-x}\right)\right) \ly+8
   \left(1-\frac{1}{1-x}\right) \ls \ly+4
   \left(1-\frac{1}{1-x}\right) \pi^2
\ibrk
+8
   \left(x-\frac{1}{1-x}+2\right) \lm^2-32
   \left(x+\left(1-\frac{1}{1-x}\right) \z3\right)-16
   \left(x-\frac{2}{1-x}+4\right) \lm
\ibrk
-8 \left(-2
   x-\frac{9}{1-x}+9\right) \ls
+16
   \left(x-\frac{1}{1-x}+2\right) \lm \ls-16
   \left(1-\frac{1}{1-x}\right) \s12
\right]
\right \} \nonumber \\ &&
+ \epsilon \,i \pi \left \{
\frac{1}{m_s^2}
\left[
-16 (1-x) x
\right]
+
\frac{1}{m_s}
\left[
-48
\right]
+
\left[
8 \left(1-\frac{1}{1-x}\right) \ly^2+16
   \left(1-\frac{2}{1-x}\right) \ly
\ibrk
+16
   \left(1-\frac{1}{1-x}\right) \ls \ly
+16
   \left(-x-\frac{5}{1-x}+4\right)+16 \left(1-\frac{1}{1-x}\right)
   \li2
\ibrk
-16 \left(x-\frac{1}{1-x}+2\right) \lm+8
   \left(1-\frac{1}{1-x}\right) \ls
\right]
\right \} \nonumber \\ &&
+ \epsilon^2 
\left \{
\frac{1}{m_s^2}
\left[
\frac{2}{15} (1-x) x \pi^4+\frac{28}{3} (1-x) x
   \pi^2-8 (1-x) x \ls^2-64 (1-x) x+32 (1-x) x
   \ls
\ibrk
+\z3 (12 (1-x) x-8 (1-x) x \ls)
\right]
+
\frac{1}{m_s}
\left[
\frac{2 \pi^4}{5}+28 \pi^2-24
   \ls^2+\z3 (4-24 \ls)+32
   \ls-64
\right]
\brk
+
\left[
-\frac{8}{15} \left(1-\frac{1}{1-x}\right) \pi^4-\frac{2}{3}
   \left(-14 x-\frac{69}{1-x}+57\right) \pi^2-8
   \left(1-\frac{1}{1-x}\right) \li2 \pi^2
\ibrk
-2
   \left(1-\frac{1}{1-x}\right) \ly^4-\frac{8}{3}
   \left(x-\frac{1}{1-x}+2\right) \lm^3-\frac{4}{3}
   \left(5-\frac{9}{1-x}\right) \ly^3-\frac{16}{3}
   \left(1-\frac{1}{1-x}\right) \ls \ly^3
\ibrk
+8
   \left(x-\frac{2}{1-x}+4\right) \lm^2+4 \left(-2
   x-\frac{9}{1-x}+9\right) \ls^2-8
   \left(x-\frac{1}{1-x}+2\right) \lm \ls^2
\ibrk
-4
   \left(1-\frac{1}{1-x}\right) \ls^2
   \ly^2+\left(12 \left(1+\frac{1}{1-x}\right)-\frac{10}{3}
   \left(1-\frac{1}{1-x}\right) \pi^2\right)
   \ly^2-4 \left(3-\frac{5}{1-x}\right) \ls
   \ly^2
\ibrk
+8 \left(-8 x+\left(2-\frac{2}{1-x}\right)
   \z3-\frac{9}{1-x}+9\right)+\left(\frac{28}{3}
   \left(x-\frac{1}{1-x}+2\right) \pi^2-16 \left(2
   x-\frac{3}{1-x}+6\right)\right) \lm
\ibrk
-8
   \left(x-\frac{1}{1-x}+2\right) \lm^2
   \ls+\left(-4 \left(1-\frac{1}{1-x}\right)
   \pi^2+32 x+\left(32-\frac{32}{1-x}\right) \z3\right)
   \ls
\ibrk
+16 \left(x-\frac{2}{1-x}+4\right) \lm
   \ls-4 \left(1-\frac{1}{1-x}\right) \ls^2
   \ly+\left(-\frac{2}{3} \left(11-\frac{23}{1-x}\right)
   \pi^2-24 \left(1+\frac{1}{1-x}\right)\right)
   \ly
\ibrk
+\left(8 \left(1+\frac{1}{1-x}\right)-8
   \left(1-\frac{1}{1-x}\right) \pi^2\right) \ls
   \ly+16 \left(1-\frac{2}{1-x}\right) \s12+16
   \left(1-\frac{1}{1-x}\right) \ls \s12
\ibrk
+16
   \left(1-\frac{1}{1-x}\right) \ly \s12+16
   \left(1-\frac{1}{1-x}\right) \s13-16
   \left(1-\frac{1}{1-x}\right) \s22
\right]
\right \} \nonumber \\ &&
+ \epsilon^2 \, i \pi
\left \{
\frac{1}{m_s^2}
\left[
-32 (1-x) x+8 (1-x) \z3 x+16 (1-x) \ls x
\right]
+
\frac{1}{m_s}
\left[
24 \z3+48 \ls-32
\right]
\brk
+
\left[
-\frac{8}{3} \left(1-\frac{1}{1-x}\right) \ly^3-8
   \left(1-\frac{2}{1-x}\right) \ly^2-8
   \left(1-\frac{1}{1-x}\right) \ls \ly^2-8
   \left(1-\frac{1}{1-x}\right) \ls^2 \ly
\ibrk
+\left(4
   \left(1-\frac{1}{1-x}\right) \pi^2+16
   \left(1+\frac{1}{1-x}\right)\right) \ly-16
   \left(1-\frac{1}{1-x}\right) \li2 \ly-16
   \left(1-\frac{2}{1-x}\right) \ls \ly
\ibrk
+2
   \left(1-\frac{1}{1-x}\right) \pi^2+8
   \left(x-\frac{1}{1-x}+2\right) \lm^2-4
   \left(1-\frac{1}{1-x}\right) \ls^2
\ibrk
-8 \left(4
   x+\left(4-\frac{4}{1-x}\right) \z3+\frac{3}{1-x}+3\right)-16
   \left(1-\frac{2}{1-x}\right) \li2+16
   \left(1-\frac{1}{1-x}\right) \li3
\ibrk
-16
   \left(x-\frac{2}{1-x}+4\right) \lm-16
   \left(-x-\frac{5}{1-x}+4\right) \ls-16
   \left(1-\frac{1}{1-x}\right) \li2 \ls
\ibrk
+16
   \left(x-\frac{1}{1-x}+2\right) \lm \ls-16
   \left(1-\frac{1}{1-x}\right) \s12
\right]
\right \} \, ,
\eea

\bea
{\mathcal J^{(0 \times 1)}_2} &=& 
\left \{ 
\frac{1}{m_s^2}
\left[
-16 (1-x) x
\right]
+
\frac{1}{m_s}
\left[
-16 \left(2 x^2-2 x+3\right)
\right]
+
\left[
4 (x-2) \ly^2+4 \left(x-\frac{2}{1-x}+2\right)
   \ly
\ibrk
+4 \left(-17 x-\frac{2}{1-x}+26\right)-8
   \left(x-\frac{1}{1-x}+2\right) \lm
\right]
\right \} \nonumber \\ && 
+ i \pi
\left \{
\left[
4 \left(x-\frac{2}{1-x}+2\right)+8 (x-2) \ly
\right]
\right \} \nonumber \\ &&
+\epsilon 
\left \{
\frac{1}{m_s^2}
\left[
-32 (1-x) x+8 (1-x) \z3 x+16 (1-x) \ls x
\right]
+
\frac{1}{m_s}
\left[
-32 \left(2 x^2-2 x+1\right)
\ibrk
+8 \left(2 x^2-2 x+3\right) \z3+16 \left(2
   x^2-2 x+3\right) \ls
\right]
+
\left[
-\frac{8}{3} (x-2) \ly^3+2 \left(\frac{2}{1-x}-5 x\right)
   \ly^2
\ibrk
-4 (x-2) \ls \ly^2
+\left(-4
   (x-2) \pi^2-8 \left(\frac{1}{1-x}-2 x\right)\right)
   \ly-4 \left(x-\frac{2}{1-x}+2\right) \ls
   \ly
\ibrk
-2 \left(x-\frac{2}{1-x}+2\right) \pi^2+4
   \left(x-\frac{1}{1-x}+2\right) \lm^2+4 \left(-17 x+(8 x-12)
   \z3-\frac{2}{1-x}+18\right)
\ibrk
-8 \left(x-\frac{2}{1-x}+4\right)
   \lm-4 \left(-17 x-\frac{2}{1-x}+26\right) \ls+8
   \left(x-\frac{1}{1-x}+2\right) \lm \ls+8 (x-2)
   \s12
\right]
 \right \} \nonumber \\ &&
+ \epsilon \, i \pi
\left \{
\frac{1}{m_s^2}
\left[
-16 (1-x) x
\right]
+
\frac{1}{m_s}
\left[
-16 \left(2 x^2-2 x+3\right)
\right]
+
\left[
-4 (x-2) \ly^2-8 (2 x-1) \ly
\ibrk
-8 (x-2)
   \ls \ly+4 \left(-13 x-\frac{4}{1-x}+26\right)-8
   (x-2) \li2-8 \left(x-\frac{1}{1-x}+2\right)
   \lm
\ibrk
-4 \left(x-\frac{2}{1-x}+2\right) \ls
\right]
\right \} \nonumber \\ &&
+ \epsilon^2 
\left \{ 
\frac{1}{m_s^2}
\left[
\frac{2}{15} (1-x) x \pi^4+\frac{28}{3} (1-x) x
   \pi^2-8 (1-x) x \ls^2-64 (1-x) x+32 (1-x) x
   \ls
\ibrk
+\z3 (12 (1-x) x-8 (1-x) x \ls)
\right]
+
\frac{1}{m_s}
\left[
\frac{2}{15} \left(2 x^2-2 x+3\right) \pi^4+\frac{28}{3}
   \left(2 x^2-2 x+3\right) \pi^2
\ibrk
-8 \left(2 x^2-2 x+3\right)
   \ls^2-64 \left(2 x^2-2 x+1\right)+32 \left(2 x^2-2 x+1\right)
   \ls+\z3 \left(4 \left(6 x^2-6 x+1\right)
\dibrk
-8 \left(2
   x^2-2 x+3\right) \ls\right)
\right]
+
\left[
\frac{4}{15} (2 x-3) \pi^4+\frac{1}{3} \left(95
   x+\frac{26}{1-x}-182\right) \pi^2
\ibrk
+4 (x-2) \li2
   \pi^2+(x-2) \ly^4-\frac{4}{3}
   \left(x-\frac{1}{1-x}+2\right) \lm^3-\frac{2}{3} \left(-9
   x+\frac{2}{1-x}+2\right) \ly^3
\ibrk
+\frac{8}{3} (x-2)
   \ls \ly^3+4 \left(x-\frac{2}{1-x}+4\right)
   \lm^2+2 \left(-17 x-\frac{2}{1-x}+26\right)
   \ls^2
\ibrk
-4 \left(x-\frac{1}{1-x}+2\right) \lm
   \ls^2+2 (x-2) \ls^2
   \ly^2
+\left(\frac{5}{3} (x-2) \pi^2-4
   \left(x-\frac{1}{1-x}+1\right)\right) \ly^2
\ibrk
-2
   \left(\frac{2}{1-x}-5 x\right) \ls \ly^2+8
   \left(-17 x+(2 x-1) \z3-\frac{2}{1-x}+18\right)+\left(\frac{14}{3}
   \left(x-\frac{1}{1-x}+2\right) \pi^2
\dibrk
-8 \left(2
   x-\frac{3}{1-x}+6\right)\right) \lm-4
   \left(x-\frac{1}{1-x}+2\right) \lm^2
   \ls+\left(2 \left(x-\frac{2}{1-x}+2\right)
   \pi^2+68 x
\dibrk
+(48-32 x) \z3+\frac{8}{1-x}-72\right)
   \ls+8 \left(x-\frac{2}{1-x}+4\right) \lm
   \ls+2 \left(x-\frac{2}{1-x}+2\right) \ls^2
   \ly
\ibrk
+\left(\frac{1}{3} \left(23 x+\frac{2}{1-x}-14\right)
   \pi^2-4 \left(-5 x+\frac{4}{1-x}+2\right)\right)
   \ly
\ibrk
+\left(4 (x-2) \pi^2+8
   \left(\frac{1}{1-x}-2 x\right)\right) \ls \ly-8
   (2 x-1) \s12-8 (x-2) \ls \s12
\ibrk
-8
   (x-2) \ly \s12-8 (x-2) \s13+8
   (x-2) \s22
\right]
\right \} \nonumber \\ &&
+ \epsilon^2 \, i \pi
\left \{
\frac{1}{m_s^2}
\left[
-32 (1-x) x+8 (1-x) \z3 x+16 (1-x) \ls x
\right]
+
\frac{1}{m_s}
\left[
-32 \left(2 x^2-2 x+1\right)
\ibrk
+8 \left(2 x^2-2 x+3\right) \z3+16 \left(2
   x^2-2 x+3\right) \ls
\right]
+
\left[
\frac{4}{3} (x-2) \ly^3+4 (2 x-1) \ly^2+4 (x-2)
   \ls \ly^2
\ibrk
+4 (x-2) \ls^2
   \ly+\left(8 (x-1)-2 (x-2) \pi^2\right)
   \ly+8 (x-2) \li2 \ly+8 (2 x-1)
   \ls \ly
\ibrk
+\left(-x+\frac{2}{1-x}-2\right)
   \pi^2+4 \left(x-\frac{1}{1-x}+2\right) \lm^2+2
   \left(x-\frac{2}{1-x}+2\right) \ls^2
\ibrk
+8 \left(-6 x+(4 x-6)
   \z3-\frac{3}{1-x}+8\right)+8 (2 x-1) \li2-8 (x-2)
   \li3
\ibrk
-8 \left(x-\frac{2}{1-x}+4\right) \lm-4
   \left(-13 x-\frac{4}{1-x}+26\right) \ls+8 (x-2)
   \li2 \ls
\ibrk
+8 \left(x-\frac{1}{1-x}+2\right)
   \lm \ls+8 (x-2) \s12
\right]
\right \} \, ,
\eea

\bea
{\mathcal J^{(0 \times 1)}_3} &=& 
\left \{
\frac{1}{m_s^2}
\left[
-8 (1-x) x
\right]
+
\frac{1}{m_s}
\left[
-8 \left(4 x^2-4 x+3\right)
\right]
+
\left[
96 \left(x^2-x+1\right)
\right]
\right \} \nonumber \\ &&
+\epsilon 
\left \{
\frac{1}{m_s^2}
\left[
-16 (1-x) x+4 (1-x) \z3 x+8 (1-x) \ls x
\right]
+
\frac{1}{m_s}
\left[
-16 \left(4 x^2-4 x+1\right)
\ibrk
+4 \left(4 x^2-4 x+3\right) \z3+8 \left(4
   x^2-4 x+3\right) \ls
\right]
+
\left[
-16 \left(-8 x^2+8 x+\left(3 x^2-3 x+3\right) \z3-4\right)
\ibrk
-96
   \left(x^2-x+1\right) \ls
\right]
\right \} \nonumber \\ &&
+\epsilon \, i \pi
\left \{
\frac{1}{m_s^2}
\left[
-8 (1-x) x
\right]
+
\frac{1}{m_s}
\left[
-8 \left(4 x^2-4 x+3\right)
\right]
+
\left[
96 \left(x^2-x+1\right)
\right]
\right \} \nonumber \\ &&
+ \epsilon^2 
\left \{ 
\frac{1}{m_s^2}
\left[
\frac{1}{15} (1-x) x \pi^4+\frac{14}{3} (1-x) x
   \pi^2-4 (1-x) x \ls^2-32 (1-x) x+16 (1-x) x
   \ls
\ibrk
+\z3 (6 (1-x) x-4 (1-x) x \ls)
\right]
+
\frac{1}{m_s}
\left[
\frac{1}{15} \left(4 x^2-4 x+3\right) \pi^4+\frac{14}{3}
   \left(4 x^2-4 x+3\right) \pi^2
\ibrk
-4 \left(4 x^2-4 x+3\right)
   \ls^2-32 \left(4 x^2-4 x+1\right)+16 \left(4 x^2-4 x+1\right)
   \ls+\z3 \left(2 \left(12 x^2-12 x+1\right)
\dibrk
-4 \left(4
   x^2-4 x+3\right) \ls\right)
\right]
+
\left[
-\frac{4}{5} \left(x^2-x+1\right) \pi^4-56 \left(x^2-x+1\right)
   \pi^2+48 \left(x^2-x+1\right) \ls^2
\ibrk
-8
   \left(-32 x^2+32 x+\left(5 x^2-5 x+1\right) \z3-16\right)+16 \left(-8
   x^2+8 x+\left(3 x^2-3 x+3\right) \z3-4\right) \ls
\right]
\right \} \nonumber \\ &&
+ \epsilon^2 \, i \pi 
\left \{ 
\frac{1}{m_s^2}
\left[
-16 (1-x) x+4 (1-x) \z3 x+8 (1-x) \ls x
\right]
+
\frac{1}{m_s}
\left[
-16 \left(4 x^2-4 x+1\right)
\ibrk
+4 \left(4 x^2-4 x+3\right) \z3+8 \left(4
   x^2-4 x+3\right) \ls
\right]
+
\left[
-16 \left(-8 x^2+8 x+\left(3 x^2-3 x+3\right) \z3-4\right)
\ibrk
-96
   \left(x^2-x+1\right) \ls
\right]
\right \} \, .
\eea
}
%
%
%
%

\end{document}